\documentclass[fleqn,usenatbib]{mnras}

\usepackage{newtxtext,newtxmath}

\usepackage[T1]{fontenc}
\usepackage{ae,aecompl}

\usepackage{graphicx}
\usepackage{amsmath}
\usepackage{amssymb}
\usepackage{color,ulem}

\title[Testing galaxy association methods]{
Siblings, friends and acquaintances:\\
Testing galaxy association methods\thanks{}}
\author[Caso \& Vega--Mart\'inez]
{Caso J.P.$^{1,2}$\thanks{E-mails:jpcaso@fcaglp.unlp.edu.ar (JPC); cnvega@dfuls.cl (CVM)}, Vega--Mart\'inez C.A.$^{3,4}$\\ 
$^1$Facultad de Ciencias Astron\'omicas y Geof\'isicas de la Universidad Nacional de     
La Plata, and Instituto de Astrof\'isica de La Plata \\
 (CCT La Plata -- CONICET, UNLP), Paseo del Bosque S/N, B1900FWA La Plata, Argentina\\   
$^2$Consejo Nacional de Investigaciones Cient\'ificas y T\'ecnicas, Rivadavia 1917, 
C1033AAJ Ciudad Aut\'onoma de Buenos Aires, Argentina\\
$^3$Instituto de Investigaci\'on Multidisciplinar en Ciencia y Tecnolog\'ia, 
Universidad de La Serena, Ra\'ul Bitr\'an 1305, La Serena, Chile\\
$^4$Departamento de F\'isica y Astronom\'ia, Universidad de La Serena, 
Av. Juan Cisternas 1200 Norte, La Serena, Chile
}

\date{Accepted XXX. Received YYY; in original form ZZZ}

\pubyear{2019}  

\begin{document}
\label{firstpage}
\pagerange{\pageref{firstpage}--\pageref{lastpage}}
\maketitle

\begin{abstract}
In order to constraint the limitations of association methods 
applied to galaxy surveys, we analysed the
catalogue of halos at $z=0$ of a cosmological simulation,
trying to reproduce the limitations that an observational
survey deal with. We focused in the percolation method, 
usually called Friends of Friends method, commonly used
in literature. The analysis was carried on the dark matter 
cosmological simulation MDPL2, from the Multidark project. 
Results point to a large fraction of contaminants for massive 
halos in high density environments. Thresholds in the
association parameters and the subsequent analysis of 
observational properties can mitigate the occurrence of fake 
positives. The use of tests for substructures can also
be efficient in particular cases. 
\end{abstract}


\section{Introduction}
\label{intro}
The evolution of the galaxies and their current observed 
properties are a consequence of the combined effects of
self-regulated internal processes and external ones related 
with the environment where they lie. For instance, it is well established 
in literature that high density environments, like clusters of galaxies, 
present a larger fraction of early-type galaxies than the field  
\citep[e.g.][]{bal06,ras12,fin18}, but the processes ruling this quenching 
are not clearly established and they have been subject of study in the 
last years \citep{kaw17}. 
Several studies also point to differences in the properties of early-type
galaxies depending on their environment, with field ellipticals 
having lower metallicities \citep{nie10,lac16}, lighter halos \citep{men09,sal12,ric15} and a larger proportion of intermediate-age 
mergers \citep{her08,tal09,hir13} than those located in clusters. 
These topics demand a straight comparison between properties obtained
from numerical simulations and observational surveys.
This becomes more relevant with the new
generation of numerical simulations \citep{vog14,sch15,kne18}, and the 
currently available galaxy surveys like Sloan Digital Sky Survey 
\citep[SDSS][]{yor00} or the 2MASS Extended Sources Calatogue \citep{skr06}, and 
upcoming projects like the Large Synoptic Survey Telescope 
\citep[LSST][]{ive08}.

Therefore, an accurate classification of the environments where the 
galaxies reside becomes
a key point to understand its role in galaxy evolution. Several attempts  
to identify nearby groups and clusters of galaxies 
have been made in the past decades \citep[e.g.][]{dev75,pat79},
improving the accuracy of the methods through
the development of algorithms 
based on objective methods like the linking-length percolation
\citep[][]{huc82}, or the hierarchical 
clustering
\citep{mat78,tul88}. These methods and similar 
ones derived from them have been widely applied 
to successive observational surveys that
increased the
number of galaxies in the local volume with 
more precise radial velocities measurements
\citep[e.g.][]{gar93,cro07,mak11}.  
Global properties have proven to be consistent
between several studies. For instance, \citet{gar93} applied a combination
of percolation and hierarchical methods, finding that $\approx 42\%$ of 
the galaxies in the sample were clustered in groups of three or more members. 
Similar results were obtained by \citet{mak11} using the percolation
method ($\approx 45\%$). Nonetheless,  
\citet{cro07} applied the percolation method
assuming two different sets of linking-length parameters, 
namely two density contrast levels, and the 
results assuming the low density contrast
parameters double those for the high density ones ($\approx 59\%$ and 
$\approx36\%$, respectively). This latter result points to the importance 
of the selection of parameters 
and the physical motivation for preferring certain values,
which might lead to differing interpretations for the same group/cluster of
galaxies \citep[e.g.][]{cas15a,hes15}. A more complex discussion emerges 
when we also consider the surroundings of clusters of galaxies as 
possible infalling regions \citep[e.g.][]{kim14}, but this approach needs 
an extensive discussion about the definition of a cluster of galaxies, 
which exceeds the goals of the present study.

In several catalogues of nearby groups of galaxies based on redshift-space 
extragalactic databases authors have also analysed mock galaxy
catalogues obtained from N-body simulations to test their results 
\citep[e.g.][]{ber06,rob11}.
In many cases, these analysis were included to determine 
the completeness of the group catalogues and/or the reliability of their 
detected members, but restricting their results by using specific 
sets of linking-length parameters and observational databases in 
each case.
\citet{eke04a} applied a percolation method to the Two-Degree Field 
Galaxy Redshift Survey (2DFGRS), to test how the velocity 
dispersion and group sizes derived for the 2DFGRS are affected by 
variations in the algorithm parameters using a mock catalogue, 
and this analysis was extended to measure the impact of variations in the
luminosity function \citep{eke04b}.
\citet{dua14} analysed the percolation method
accuracy for different sets of linking-length parameters for a SDSS-like 
mock catalogue, created from the Millennium-II simulation \citep{boy09} 
and a semi-analytic model. Besides the completeness and completeness 
percentages, they obtained global values of fragmentation and merging for 
the true groups of galaxies, assuming luminosity and distance completeness 
with maximum redshift ranging from $z\approx0.04$ to $z\approx0.13$ 
for different luminosity limits (derived from the SDSS completeness). 
A similar approach was taken by \citet{woj18}, 
comparing several association methods from recent studies, including 
percolation ones. They focused on the accuracy of dynamical mass 
estimations of galaxy clusters, using mock clusters from the Bolshoi 
cosmological simulation \citep{kly11} with masses above 
$10^{13}\,{\rm M_{\odot}}$. They found that all methods overestimate 
cluster masses when applied to contaminated samples, 
and underestimated them when the sample is incomplete, 
concluding that this might be the main source of the scatter in 
the mass scaling relation.
Recently, \citet{tem18} used the Multidark simulation \citep{kly16} 
to test contamination and completeness percentages for their galaxy 
association algorithm, based on Bayesian statistics. All these studies
present relevant results in the testing of the accuracy of association 
methods, but they are focused on particular observational properties
or observing catalogues. Hence, it is worth to perform an independent 
test of the association methods to understand their potential strengths, 
and how the limitations of observational astronomy can influence the 
measured properties of galaxy groups derived from the classification
through the incidence of fake positives and/or lost members in the 
association.

In this work we analyse the percolation method by taking advantage 
of a large sample of galaxy halos extracted from a high resolution 
cosmological dark matter simulation to model the available data of 
observational surveys and test the method, focusing on the change 
in observational properties in the nearby Universe and possible 
constraints to improve the results.
The paper is organised as follows. In Section~\ref{sec.sample} we 
describe the analysed simulation and the methods applied to assign 
magnitudes, velocities uncertainties and calculate the projected 
distances to build our simulated galaxy catalogue. 
Section~\ref{permet} describes the analysed percolation method
applied to the galaxy catalogue. Section~\ref{sec.results} shows the 
criteria to select the linking length parameters, and the analysis 
of the results from the method, including possible constraints to
improve the accuracy of the method. Finally, Section\,5 
summarises the results achieved in this work.

\section[]{The sample}
\label{sec.sample}
We analyse the MDPL2 cosmological dark matter simulation, which 
is part of the Multidark project \citep{kly16}, and is publicly 
available through the official database of the project%
\footnote{\url{https://www.cosmosim.org/}}. This simulation consists 
in a periodic cubic volume of $1\,{\rm h^{-1}\,Gpc}$ of size length, 
filled with $3840^3$ particles with mass of $1.51\times10^9\,{\rm h^{-1}\,M_\odot}$ and
it considers the cosmological parameters of \citet{pla13}. 

From the simulation we select the complete
available catalogue of dark matter halos detected using   
\textsc{Rockstar} halo finder \citep{beh13}, specifically the ones 
corresponding to the local Universe ($z=0$). 
The catalogue is composed by
main host halos found over the background density 
and satellite halos (or subhalos) lying within another halos. 
We consider each one of these structures as host of a unique galaxy, so
the main halos become hosts of the central galaxies of each
system, and the satellite halos are the hosts of the satellite galaxies.
From each halo we extracted from the catalogue its position, velocity,
host/satellite relationships and the mass and virial radius
obtained considering a constant factor $\Delta=200$ with respect to the 
critical density of the Universe.
Using this mass definition we restrict our selection to the halos 
with virial mass larger than $10^{11}\,\textrm{M}_\odot$.
In order to obtain a collection of samples similar to the nearby Universe, 
we divided the simulation in smaller cubic volumes with $100\,{\rm h^{-1}\,Mpc}$
of side length, including an overlapped envelope of 
$10\,{\rm h^{-1}\,Mpc}$ of width to avoid biases at the edges of the volumes.
Therefore, halos lying within these regions might be
associated as subhalos by the methods applied in each volume, 
but they cannot trigger a new association as main halos.

To avoid numerically expensive projection calculations, 
the analysis is carried on in the three possible projection planes 
following the basis of the Cartesian comoving coordinates. 
Distances in the line-of-sight were used to obtain recessional velocities,
assuming $H_0=67.77$\,km\,s$^{-1}$\,Mpc$^{-1}$.

\subsection{Magnitudes allocation}
\label{mag.alloc}
In addition to the halo properties described in the previous section, 
we assigned to each halo a luminosity in the $K$ band by using 
a simple implementation of a halo occupation distribution method 
\citep[HOD][]{val06, con06}, which assigns each luminosity 
in a non parametric way.
We simply assume a monotonic relation between the galaxy luminosities 
and the halo virial masses without making distinction between 
main hosts and satellites, by following 
\begin{equation}
   n_\textrm{g}( > L) = n_\textrm{h}(> M),
\end{equation}
where $n_\textrm{g}$ and $n_\textrm{h}$ are the number density of galaxies
and halos respectively. Whereas the halo number density is extracted
directly from the simulation, the number density of galaxies must preserve 
the galaxy luminosity function (LF) which is modelled 
by the \citet{sch76} parametric function,
\begin{equation}
        \frac{d n_\textrm{g}}{d L} =
        \frac{n_*}{L_*}\left( \frac{L}{L_*} \right)^\alpha
        \exp \left( -L / L_* \right).
\end{equation}
In our model, we took the $K$--band luminosity function 
measured by \citet{koc01} from the 2MASS survey to assign $K$ magnitudes,
which parameters are given by $K_* - 5 \log {\rm h} = -23.39\;\textrm{mag}$, 
$\alpha = -1.09$ and
$n_* = 1.16 \times 10^{-2}\,{\rm h^3\,Mpc^{-3}}$. 
Expressing the Schechter LF in terms of the magnitudes and starting 
from the bright end of the distribution, rest frame $M_K$ magnitudes were 
assigned to all the halos using a precision of 0.01 mag.
The most massive main halo in MDPL2 presents
a virial mass of $\approx 2 \times 10^{15}\,{\rm h^{-1}\,M_{\odot}}$, which is similar to 
the typical mass values derived for the Coma cluster in literature 
\citep[e.g.][]{gel99,lok03,kub07}.
Hence, we choose the luminosity of NGC\,4889, the brightest galaxy 
in the Coma cluster, as the upper limit for the HOD. 
If we assume $K=8.4$\,mag \citep{gav96} as the apparent magnitude 
and a distance of 94\,Mpc (mean value for distance estimations 
from NED\footnote{This research has made use of the 
NASA/IPAC Extragalactic Database (NED) which is operated 
by the Jet Propulsion Laboratory, California Institute of 
Technology, under contract with the National Aeronautics 
and Space Administration.}),
its absolute magnitude in the $K$ filter is $M_K= -26.5$.

It is worth noting that this method does not make any distinction between
the LFs of main hosts and satellites galaxies, introducing a bias in 
the assignation of magnitudes to the halos.
Environmental effects acting on galaxies hosted on satellite halos 
influence their properties to evolve differently than those of the central galaxies.
This produces a deviation from the assumed monotonic relation 
between the galaxy stellar mass and the halo dark matter mass.
Nonetheless, the assigned magnitudes are only involved in the 
estimation of the radial velocities uncertainties, and these are large 
enough to mask any contribution due to the simplified treatment. Then, 
no effect is expected in the results of our analysis.

\subsection{Velocities uncertainties}
\label{vel.uncert}
\begin{figure}
 \includegraphics[width=\columnwidth]{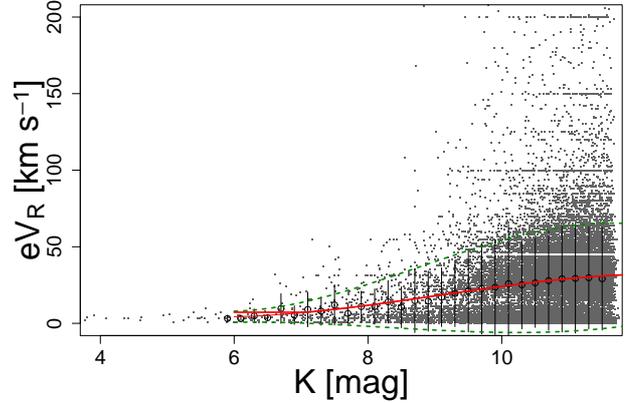}
 \caption{Uncertainties in radial velocity from the 2MASS Redshift Survey 
\citep{huc12} as a function of galaxies apparent magnitude in $K$ filter.
Open circles show the mean values in bins of 0.2\,mag, the solid red line 
represent a third-order polynomial fitted to data, while green dashed lines 
correspond to the standard deviation fit.}
 \label{6df}
\end{figure}

The uncertainties in radial velocity measurements ($eV_\textrm{R}$) are among the 
observational limitations we should take into account. In order to do this,
we analyse $V_\textrm{R}$ measurements for a sample of galaxies obtained from 
the 2MASS Redshift Survey \citep{huc12}, based on the Extended Sources 
Catalogue from 2MASS\footnote{This publication makes use of data products 
from the Two Micron All Sky Survey, which is a joint project of the 
University of Massachusetts and the Infrared Processing and Analysis 
Center/California Institute of Technology, funded by the National Aeronautics 
and Space Administration and the National Science Foundation.} \citep{skr06}. 
The Figure~\ref{6df} shows $eV_\textrm{R}$ as a function of galaxy apparent 
magnitudes in $K$ filter. 
To obtain a phenomenological model of these measurements,  
the mean values were calculated using bins of 0.2\,mag, which are 
depicted with open circles. 
A third-order polynomial was fitted to the data. The resulting curve
is represented with the solid red line, 
while the green dashed lines show one standard deviation of the fit.

For each halo, its $K$ luminosity defines three apparent magnitudes, 
depending on the Cartesian plane chosen as projected sky plane. Hence, we 
added randomly generated uncertainties to the velocity components, assuming 
Gaussian distributions for the $eV_\textrm{R}$, with mean and standard 
deviation values are defined as functions of the $K$ apparent magnitude.

\subsection{Projected distances}
\label{proj.est}

An important limitation in observational surveys of galaxies 
is the difficulty to measure accurate distances in the 
line-of-sight. This uncertainty propagates to the projected 
distances, resulting in a source of noise for the group finder 
methods. Following the criteria adopted by \citet{cro07}, for 
each pair of halos we obtain projected distances from their 
projected angular separation ($\theta$, calculated from their 
comoving spatial coordinates) and the line-of-sight distance 
estimated from the average $V_\textrm{R}$. For the latter value 
there are two sources of uncertainty: the $eV_\textrm{R}$
modelled as it was described in the previous Section, and the 
halo peculiar velocities which might represent a significant 
fraction of the recessional velocity for nearby galaxies.

\section[]{Linking-length percolation method}
\label{permet}

This family of methods, also called Friends of Friends methods,
have been widely used for galaxy clustering in observational 
astronomy \citep[e.g.][]{huc82,gar93}, but also for halo-finding 
in dark matter $N$-body simulations \citep[e.g.][and references therein]{kne11}.
In observational astronomy, the method proceed by linking a 
particular galaxy with all its neighbours that fulfil a set of 
criteria in projected distance and radial velocity, and the 
process is repeated with all the linked objects in an iterative
process until no new galaxies are associated. The result is a 
unique system where a single member does not necessarily fulfil 
the criteria with all its partners, but it does it at least with 
a single one. In this work we follow the association algorithm 
described by \citet{cro07}. This method presents two free parameters 
corresponding to the upper limits of the linking length in both 
radial velocities ($V_\textrm{R,max}$) and projected distances
($D_\textrm{p,max}$).
To apply this method in the simulated catalogues, we consider two 
lower limits in virial mass, $10^{11}$ and $10^{12}\,\textrm{M}_\odot$, 
resulting in two different samples which are analysed independently.

In this paper, we do not go further on considerations about biases related 
to the completeness of the luminosity function for flux-limited 
surveys, neither completeness in redshift space. We assume the sample is
complete up to the limiting virial mass at the entire redshift range,
which roughly span up to $10^4$\,km\,s$^{-1}$.

\section{Results}
\label{sec.results}

\subsection{Simulation statistics}
\label{simstats}

\begin{figure*}    
  \centering
  \includegraphics[width=0.4\textwidth]{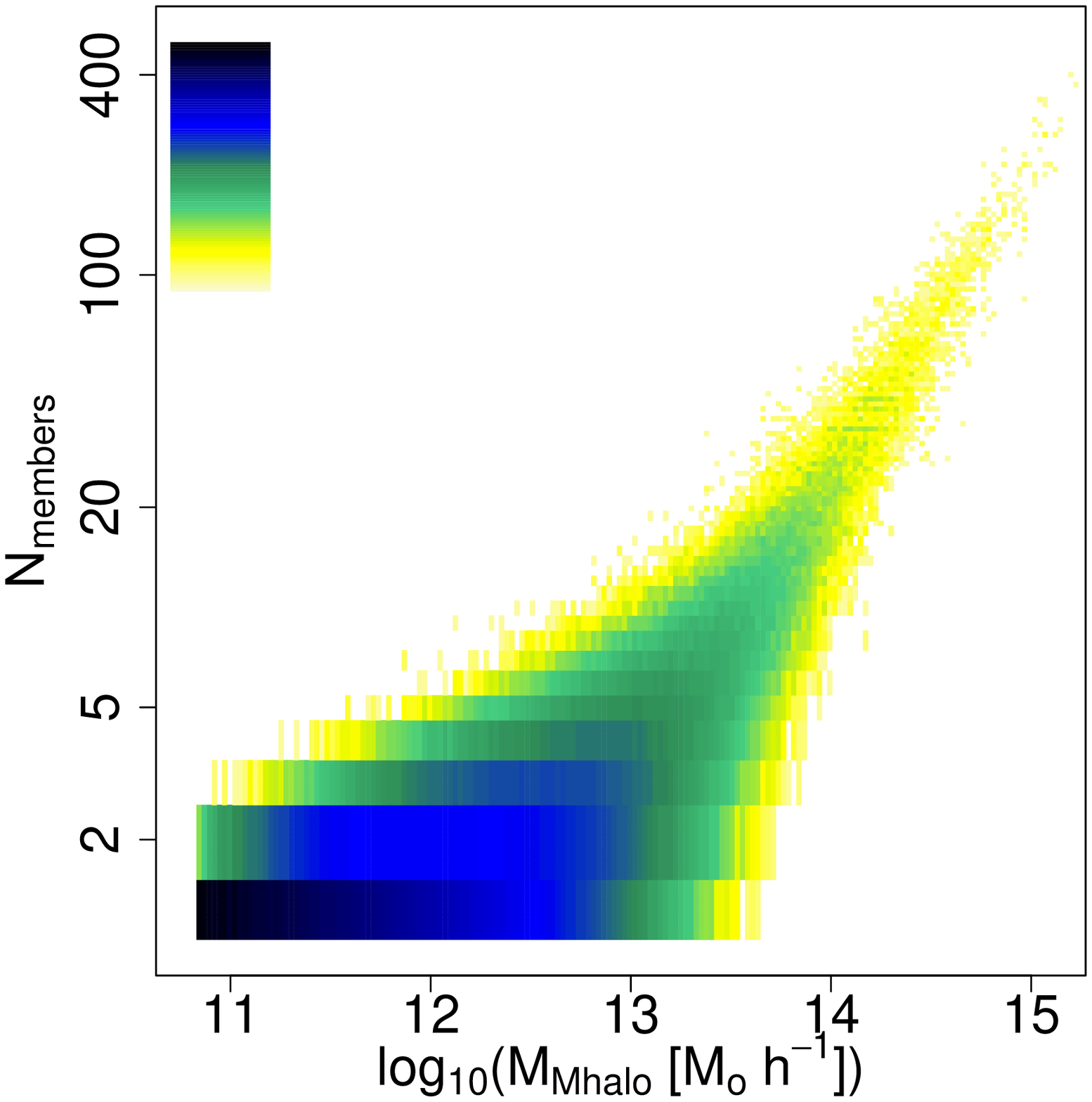} \hspace{2.5mm}   
  \includegraphics[width=0.4\textwidth]{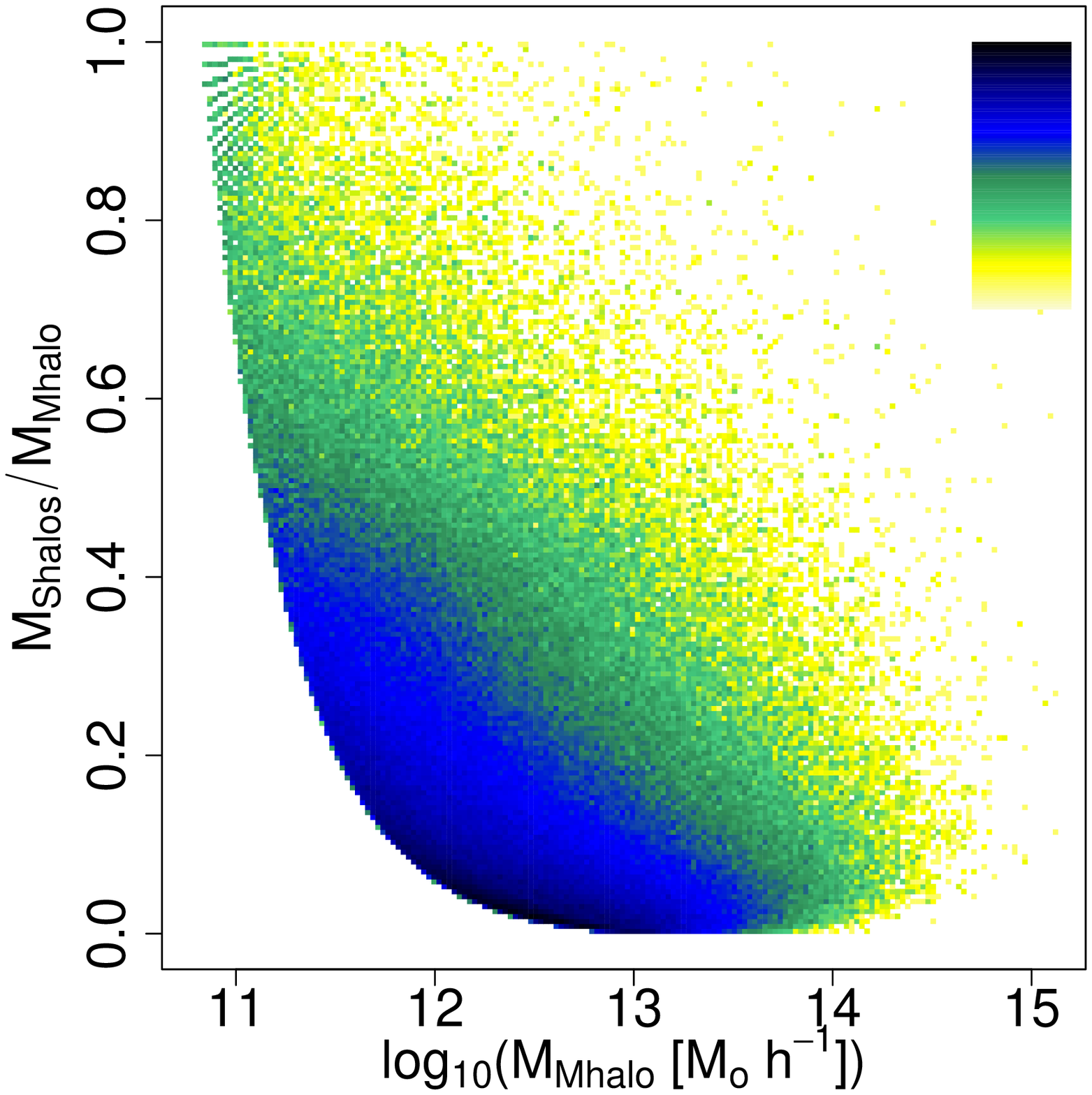} \\ \vspace{2.5mm}
  \includegraphics[width=0.4\textwidth]{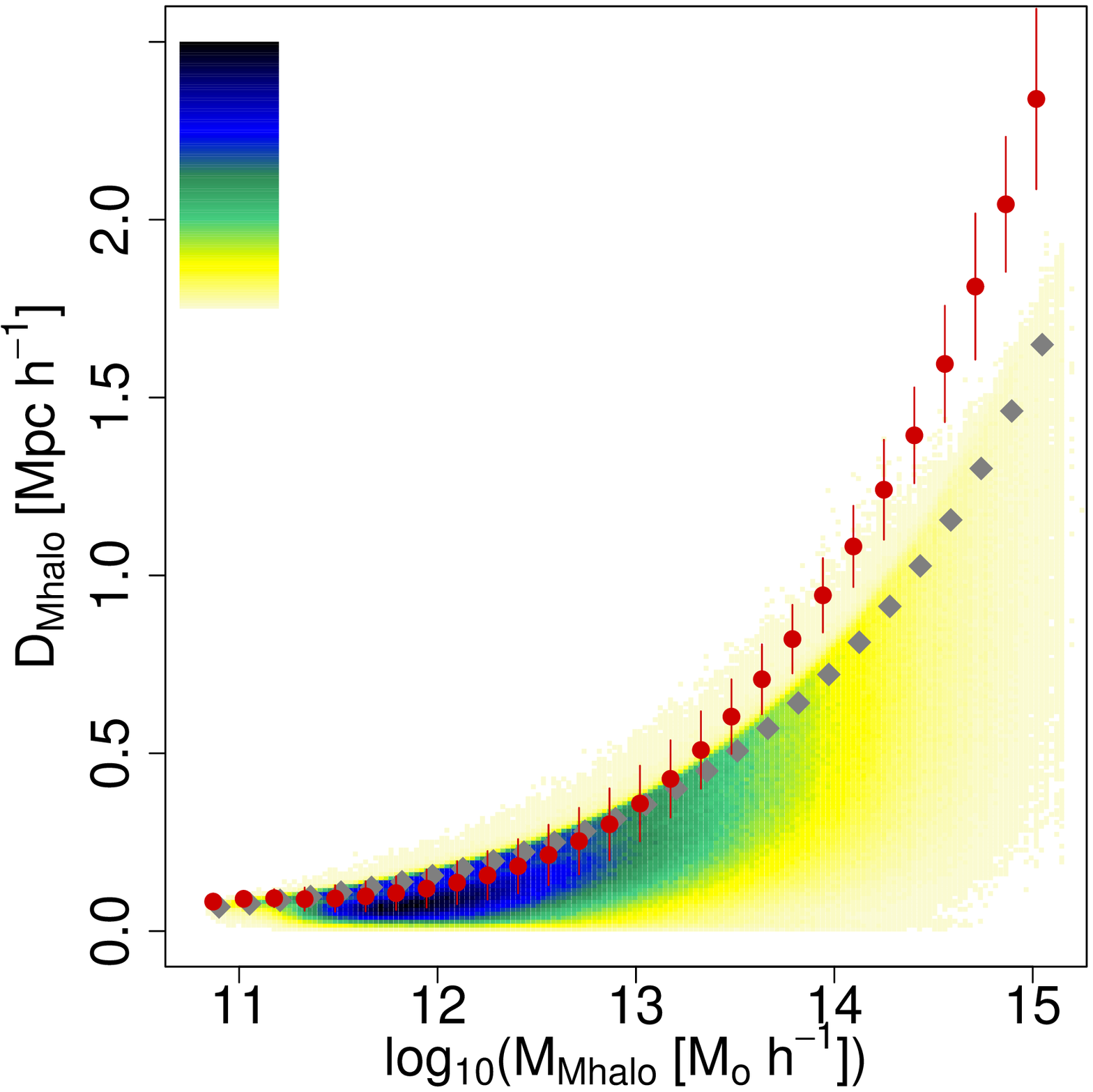}  \hspace{2.5mm}  
  \includegraphics[width=0.4\textwidth]{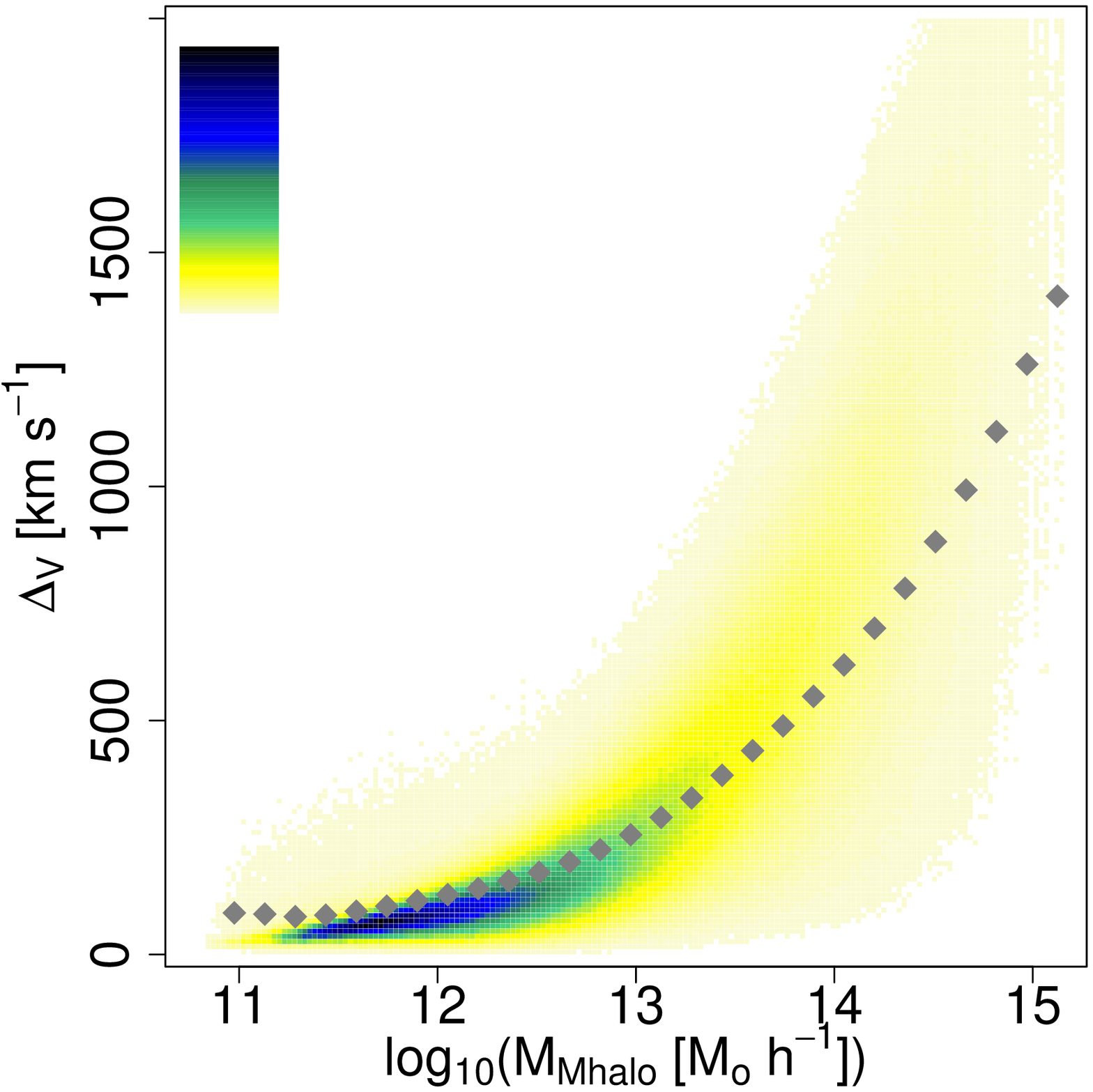}    
\caption{Distribution of different main halo
properties as a function of their virial mass. In the four panels, 
the colour gradient ranges from yellow to black according to
the value frequency. {\bf Upper-left panel:} number of components in 
main halos, 
{\bf Upper-right panel:}
fraction of mass enclosed in satellite halos to main halo mass. 
{\bf Lower-left panel:} spatial distance of satellite halos to the 
centre of the main halo. Grey diamonds indicate the mean virial radii 
($R_\textrm{vir}$), red circles represent the mean distances of the farthest 
satellite halo ($D_{\max}$) and the error-bars their dispersion
($\sigma_\textrm{D}$). 
{\bf Lower-right panel:} largest component in commovil velocities with respect 
to the main halo ($\Delta_{V,\max}$). Grey diamonds represent the mean of 
the largest component of the velocity dispersions ($\sigma_{V,\max}$) when
we assume each satellite halo as a unique galaxy.}    
\label{simul}    
\end{figure*}

As a first approach, we test some general statistical 
properties of the halos resulting from the MDPL2
simulation which are relevant in the further analysis.
These are shown in the Figure~\ref{simul}.
The upper-left panel shows the distribution of the number of
components in main halos. The y-axis is in logarithmic scale and the colour
gradient gets darker towards more populated bins. Systems with at least
ten components are typically more massive than $10^{13}\,{\rm M_{\odot}\,h^{-1}}$,
whereas those with a hundred components present virial masses around a few 
times $10^{14}\,{\rm M_{\odot}\,h^{-1}}$. The most massive single halos present 
masses around $5\times10^{13}\,{\rm M_{\odot}\,h^{-1}}$. 
The upper-right panel shows the fraction of mass contained 
in satellite halos, as a function of the main halo mass. 
While this fraction is lower than $0.5$ in the majority of 
the cases, and particularly in massive halos, in low mass 
halos it can reach larger values.

The lower-left panel in Figure~\ref{simul} shows the distribution of spatial 
distances of subhalos to the corresponding main halo, in terms of main halo
mass. Grey diamonds indicate the mean virial radii ($R_\textrm{vir}$), 
while red circles 
and their errorbars represent the mean distances of the farthest subhalo 
($D_{\max}$) and their dispersion ($\sigma_\textrm{D}$), respectively. 
This latter parameter goes beyond the $R_\textrm{vir}$ for masses larger 
than $10^{13}\,{\rm M_{\odot}\,h^{-1}}$.
For the most massive halos, $D_{\max}+\sigma_\textrm{D}$ reaches 
$\approx 1.5 R_\textrm{vir}$.

The lower-right panel in Figure~\ref{simul} shows the distribution of the 
largest component in peculiar velocities with respect to the corresponding 
main halo ($\Delta_\textrm{V,max}$) in terms of the main halo mass. Grey 
diamonds represent the mean of the largest component of the velocity 
dispersion for the system ($\sigma_\textrm{V,max}$), 
assuming each subhalo as a unique galaxy. 

For comparison purposes, we calculate for each halo a numerical proxy 
to characterise the environmental density around it. This numerical density is
defined as the number of detected halos having a physical distance lower 
than $1.75\,{\rm Mpc\,h^{-1}}$. This limiting radius 
represents the typical virial radius in the most massive main halos (see 
the third panel in Figure\,\ref{simul}). This choice points to avoid biases 
that smaller radius might introduce in the numerical density of more 
populous systems.

\subsection{Selection of linking length parameters}
\label{sec.param}

\citet{cro07} tested the variation in the percolation method when 
different parameter pairs were applied to a sample of nearby galaxy 
clusters. They found that the numbers of groups with three or more 
members was maximized  for $V_\textrm{R,max}=399$\,km\,s$^{-1}$ and 
$D_\textrm{p,max}=1.63$\,Mpc. They also derived  $V_\textrm{R,max}=350
$\,km\,s$^{-1}$ and $D_\textrm{p,max}=0.89$\,Mpc as the parameters 
corresponding to a density contrast of $\delta\rho/\rho = 80$,
typical for groups of galaxies. 
As the lower-right panel of Figure\,\ref{simul} shows,
the fraction of halos with $\Delta_\textrm{V,max}$ larger than 
these values of $V_\textrm{R,max}$ increases from $10^{12}\,{\rm M_{\odot}\,h^{-1}}$.

\begin{figure}    
\includegraphics[width=80mm]{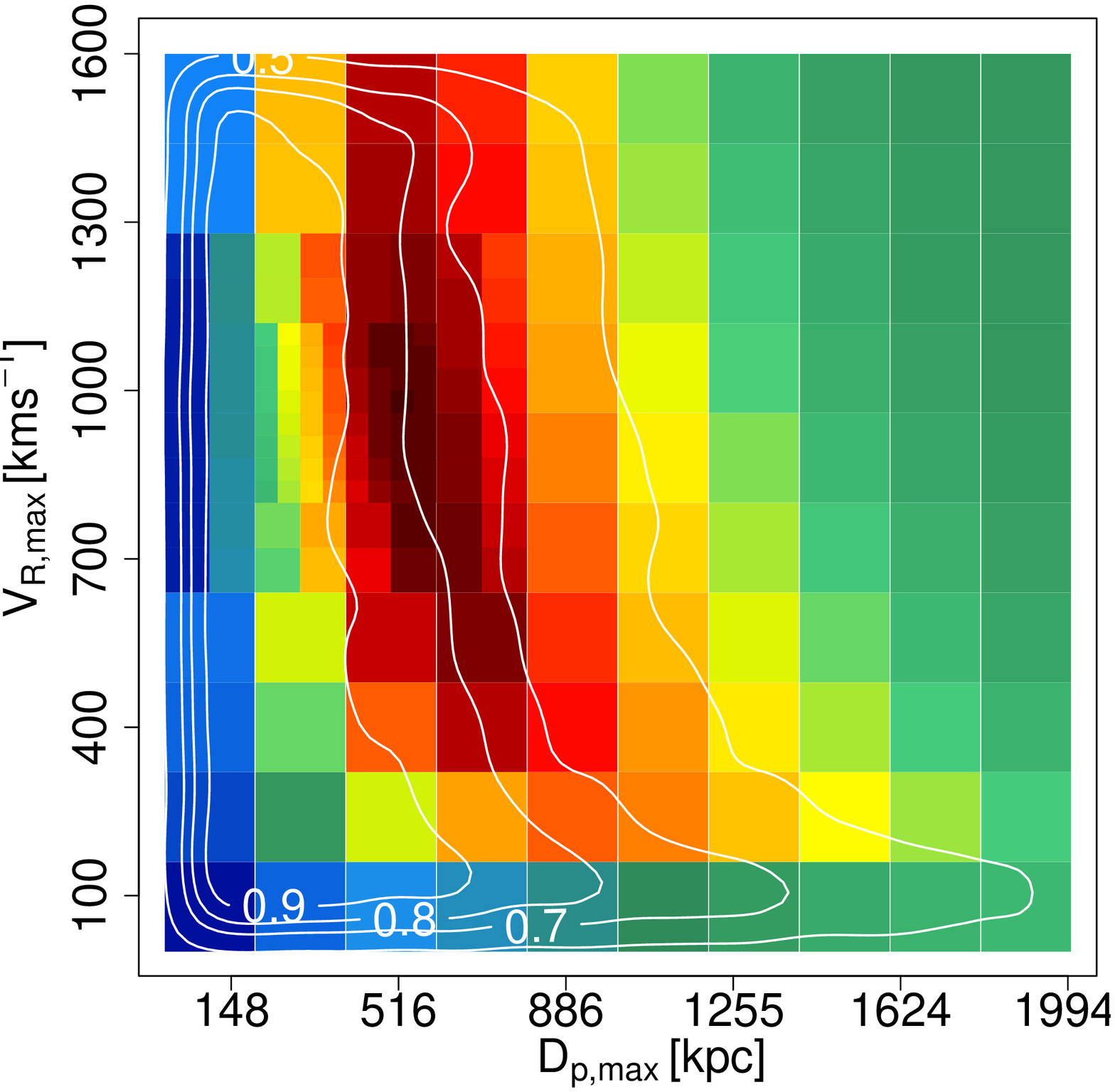}\\    
\includegraphics[width=80mm]{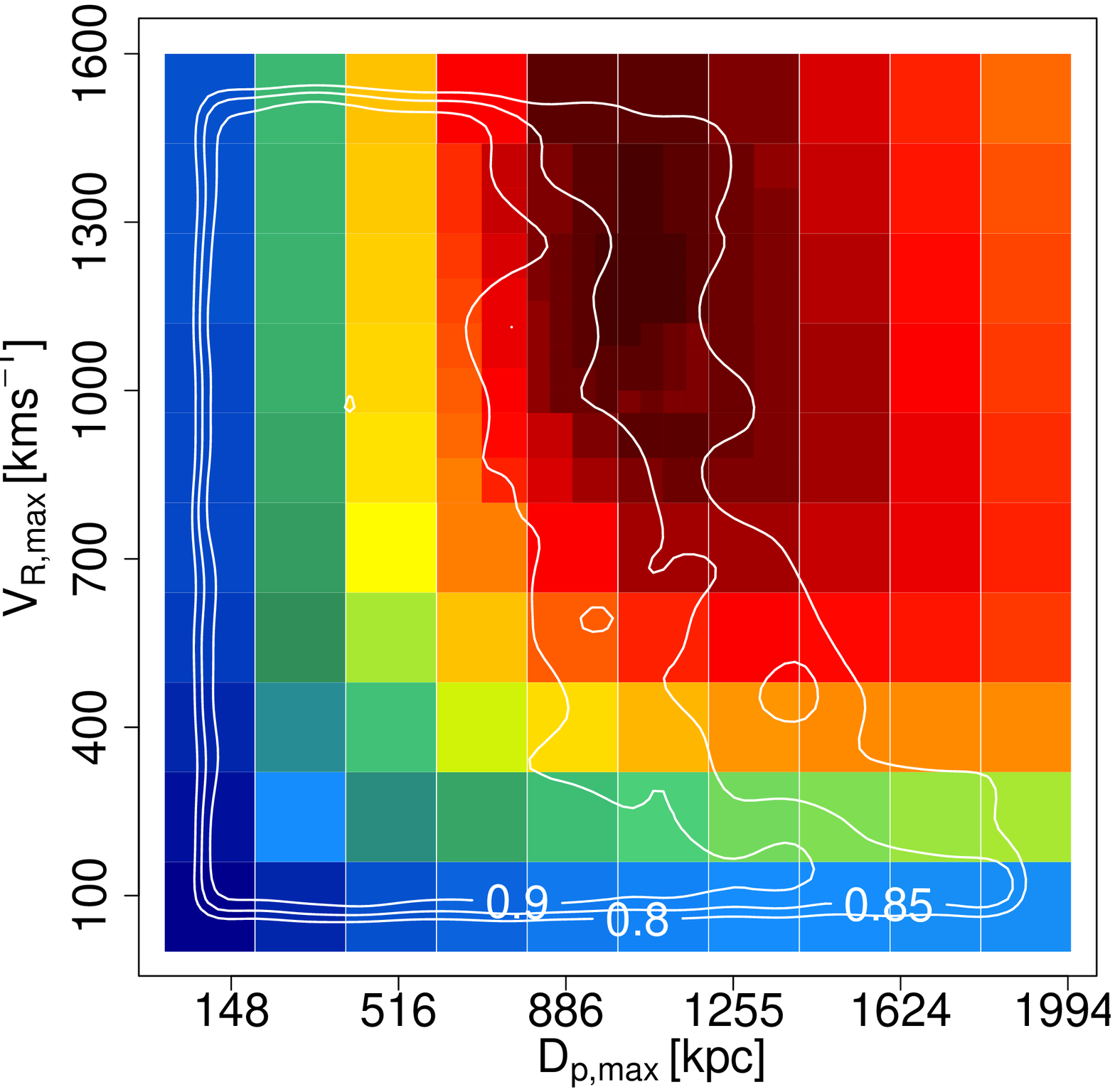}\\    
\caption{Ratio of satellite halos accurately associated to their main
halo to the total sample of satellite halos in terms of the virial mass
for different values of linking-length parameters in projected distance
$D_\textrm{p,max}$ and radial velocity $V_\textrm{R,max}$. 
The colour gradients ranges
from blue (low accuracy, $\approx 80\%$) to red (high accuracy, 
$\approx 95\%$). 
White curves indicate smoothed
contour levels when main halos instead of satellite ones are taken into
account. The upper and lower panels correspond to halos more massive 
than $10^{11}$ and $10^{12}\,\textrm{M}_\odot$, respectively.}    
\label{d0v0}    
\end{figure}

Taking advantage of the comprehensive knowledge of halos properties
in the simulation, we proceed to  
compare the performance of the free parameters of the model
by analysing the fraction of success in the allocation of satellite 
halos considering a mesh of values of the parameters. 
Assuming the association of halos in the simulation as the 
correct one, we
define the fraction of success as the ratio of satellite
halos correctly associated to their main halo to the total sample of
satellite halos, in terms of their virial mass.
In the Figure~\ref{d0v0} we analyse the behaviour of this fraction 
by applying the percolation method for associating halos 
considering different sets of parameters.
The values of these parameters ranged $D_\textrm{p,max} = 0.1- 2$\,Mpc and 
$V_\textrm{R,max} = 100 - 1600$\,km\,s$^{-1}$. Smaller steps were adopted in regions
with higher fraction of success to constrain the adopted parameters
for further analysis. The colour gradients ranges from blue (low accuracy, 
$\approx 80\%$) to red (high accuracy, $\approx 95\%$). Although this
range of accuracy seems quiet narrow, observational results derived 
under these conditions might differ significantly from the real values.
In the following Sections we will go further on this.
The upper panel in Figure\,\ref{d0v0} shows the results for the halos
more massive than $10^{11}\,\textrm{M}_\odot$. The white curves represent
the smoothed contour levels when 
only the main halos are considered in the calculation, ignoring
satellites. This comparison allows us to
emphasise the different behaviours related with changes in the linking
length parameter. As expected, main halos tend to favour small
values of $D_\textrm{p,max}$, with a weak dependence of the adopted
$V_\textrm{R,max}$.

The lower panel is analogue 
to the previous one but applying the method in the sample of halos more 
massive than $10^{12}\,\textrm{M}_\odot$. Main halos with masses under 
this limit represent $\approx 87\%$ of the number of main halos but 
only $\approx 22\%$ of the virial mass. Their absence implies an 
increase in the mean distances between main halos and translates in
wider ranges of $D_\textrm{p,max}$ for constant fractions of success.
 
In both cases, the behaviour of the fraction of success exhibit a strong 
dependence with the adopted value of $D_\textrm{p,max}$, whereas for 
$V_\textrm{R,max}$ there is a wide range of velocities which produce similar 
results. 
According to this, in both cases we manually select a definite 
set of parameters which maximises 
the fraction of success for both the main
and satellite halos. 
Considering these selected
values of the parameters and the original 
ones defined by \citet{cro07}, we define three different cases
to be analysed in more detail:
\begin{itemize}
    \item Case \textbf{(A)} with 
$D_\mathrm{p,max} = 525\,\mathrm{kpc}$ and $V_\mathrm{R,max} = 980\,\mathrm{km\;s}^{-1}$.
These give the best results for halos more massive than $10^{11}\,\textrm{M}_\odot$
    \item Case \textbf{(B)} with 
$D_\mathrm{p,max} = 975\,\mathrm{kpc}$ and $V_\mathrm{R,max} = 1100\,\mathrm{km\;s}^{-1}$.
These give the best results for halos more massive than $10^{12}\,\textrm{M}_\odot$
    \item Case \textbf{(C)} with 
$D_\mathrm{p,max} = 890\,\mathrm{kpc}$ and $V_\mathrm{R,max} = 350\,\mathrm{km\;s}^{-1}$.
These were defined by \citet{cro07}, and we apply them to the sample of halos
halos more massive than $10^{12}\,\textrm{M}_\odot$.
\end{itemize}
The choice of the more massive sample of halos for the case (C)
is motivated on comparison purposes,
considering that in the lower panel of Figure\,\ref{d0v0}
these parameters show a low fraction of success for satellite halos, 
but a higher fraction for main halos than case (B).

It is worth noting that we repeated this analysis without considering 
the added noise in the radial velocities, resulting in no appreciable
differences in the behaviour of the success fraction presented in the 
Figure\,\ref{d0v0}.

In the following we highlight the main results from the percolation method.
For simplicity, only results from a single projected plane are exposed. 
However the size of the sample avoid possible biases, which was tested from 
the comparison of the results in the three planes.

\subsection{Method success and environmental density}

\begin{figure}    
\includegraphics[width=80mm]{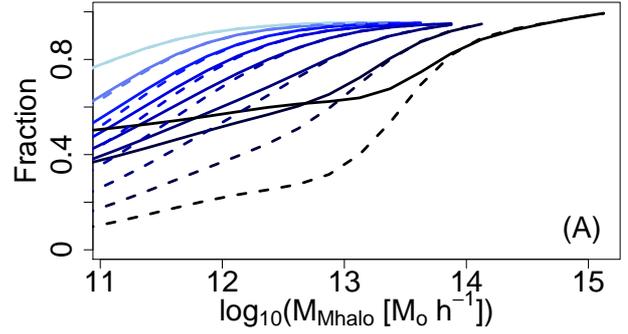}\\
\caption{Fraction of halos (solid lines) and main halos (dashed lines)
accurately classified by the method as a function of their main halo virial
mass in case (A). The sample was split in several ranges of environmental 
density which are depicted with different line colours, so that darker 
colours represent denser environments.}    
\label{frac111}    
\end{figure}

To analyse the method success on the classification 
we quantify the number of halos that are accurately classified 
by the method with respect to the total. 
For this, we define as accurately classified halos to those 
assigned by the percolation method to the same system than
the halo finder over the dark matter simulation.
According to this,
Figure~\ref{frac111} shows in solid lines the fraction of halos accurately 
classified by the method as a function of their main halo virial mass for
the selection of 
halos more massive than $10^{11}\,{\rm M_{\odot}}$. The sample was split
in percentiles of environmental density, 
so that line colours become darker towards denser environments. 
For instance, light blue corresponds to halos
presenting at most a single neighbour closer than $1.75\,{\rm Mpc h^{-1}}$,
which represents $\approx 20\%$ of the halos, whereas 
the black line
corresponds to the $5\%$ of halos located in the densest regions, typically
cluster environments.
The dashed lines correspond to main halos only, 
following the same density breaks than those applied to the general halo
population. The cut off for low density percentiles responds to the absence
of main halos more massive than a certain value.

These results indicate 
that the correct assignation of the halos has
a strong dependence on the environment, with a large 
fraction of success for halos in low density ones. 
The accuracy get worse 
for main halos in denser environments, which are associated by the method 
to more massive halos close to them. The results for the general
population of halos in these latter percentiles are noticeably better.
This improvement is due to the selection of the linking-length parameters
in Section~\ref{sec.param} was
optimised to accurately associate satellites
to their main halos.   

\begin{figure}    
\includegraphics[width=80mm]{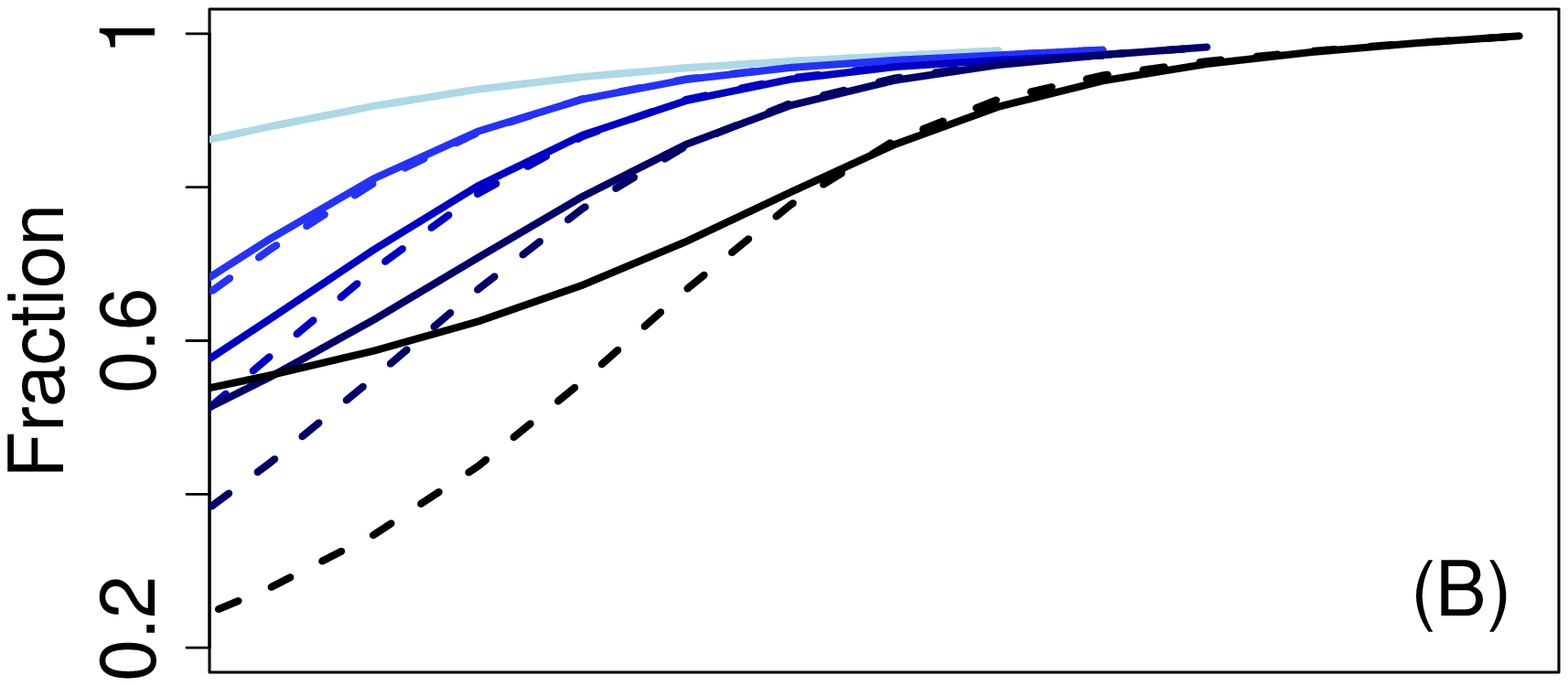}\\
\includegraphics[trim=0 0 0 2.3cm, width=80mm]{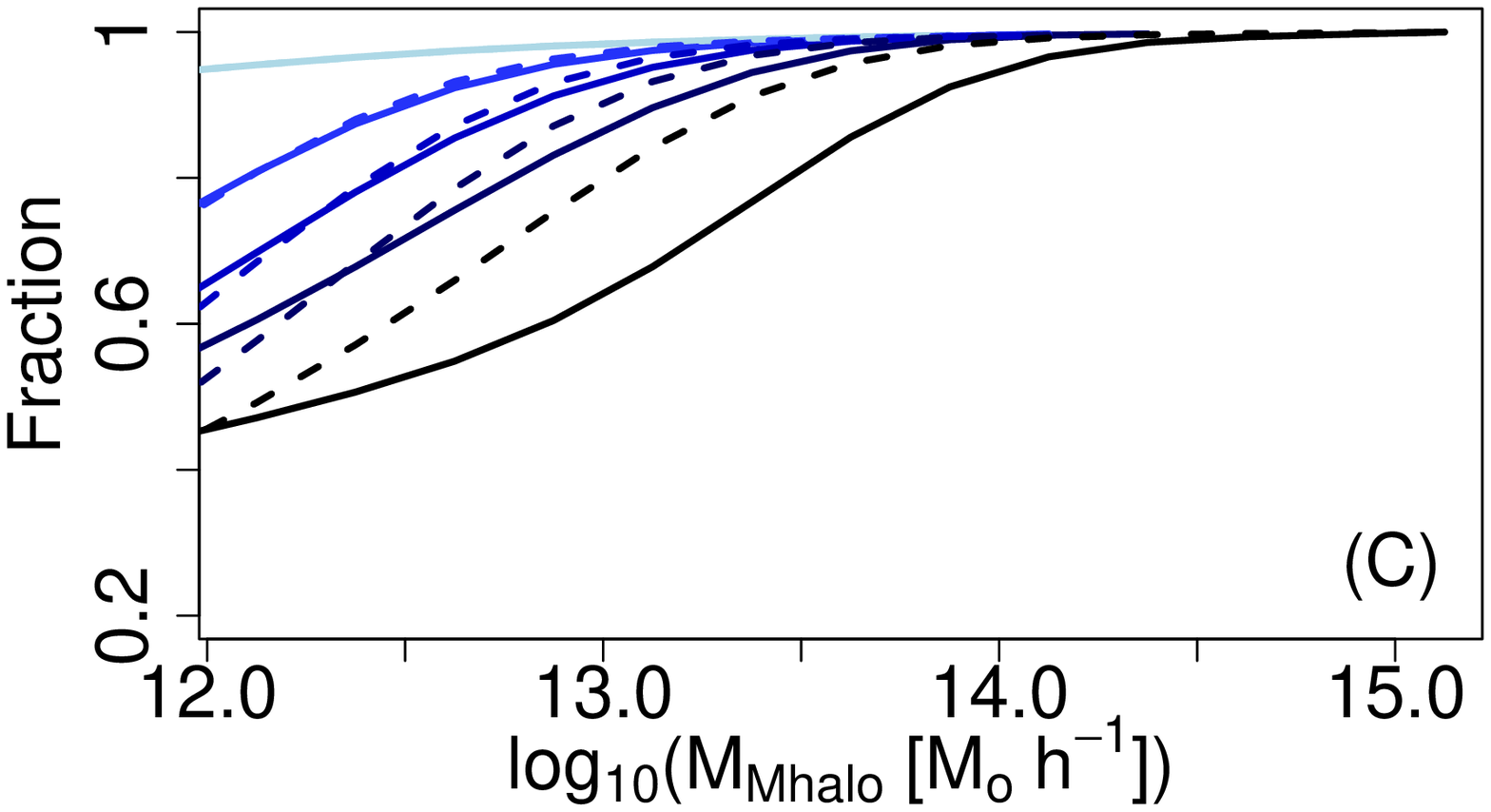}\\
\caption{Same as in Figure\,\ref{frac111}, but for halos more massive
than $10^{12}\,{\rm M_{\odot}}$ in cases (B) and (C).}    
\label{frac112}    
\end{figure}

Figure~\ref{frac112} corresponds to the set of parameters analysed for
halos more massive than $10^{12}\,{\rm M_{\odot}}$. The behaviour of the
method is analogue to the case (A) in Figure~\ref{frac111}, being less 
accurate for low mass main halos in denser environments. When we focus 
in main halos samples only, the results obtained in cases (B) and (C) 
differs significantly, as expected from their parameters chosen in 
Section~\ref{sec.param}.

\begin{figure*}    
\includegraphics[clip, width=62.5mm]{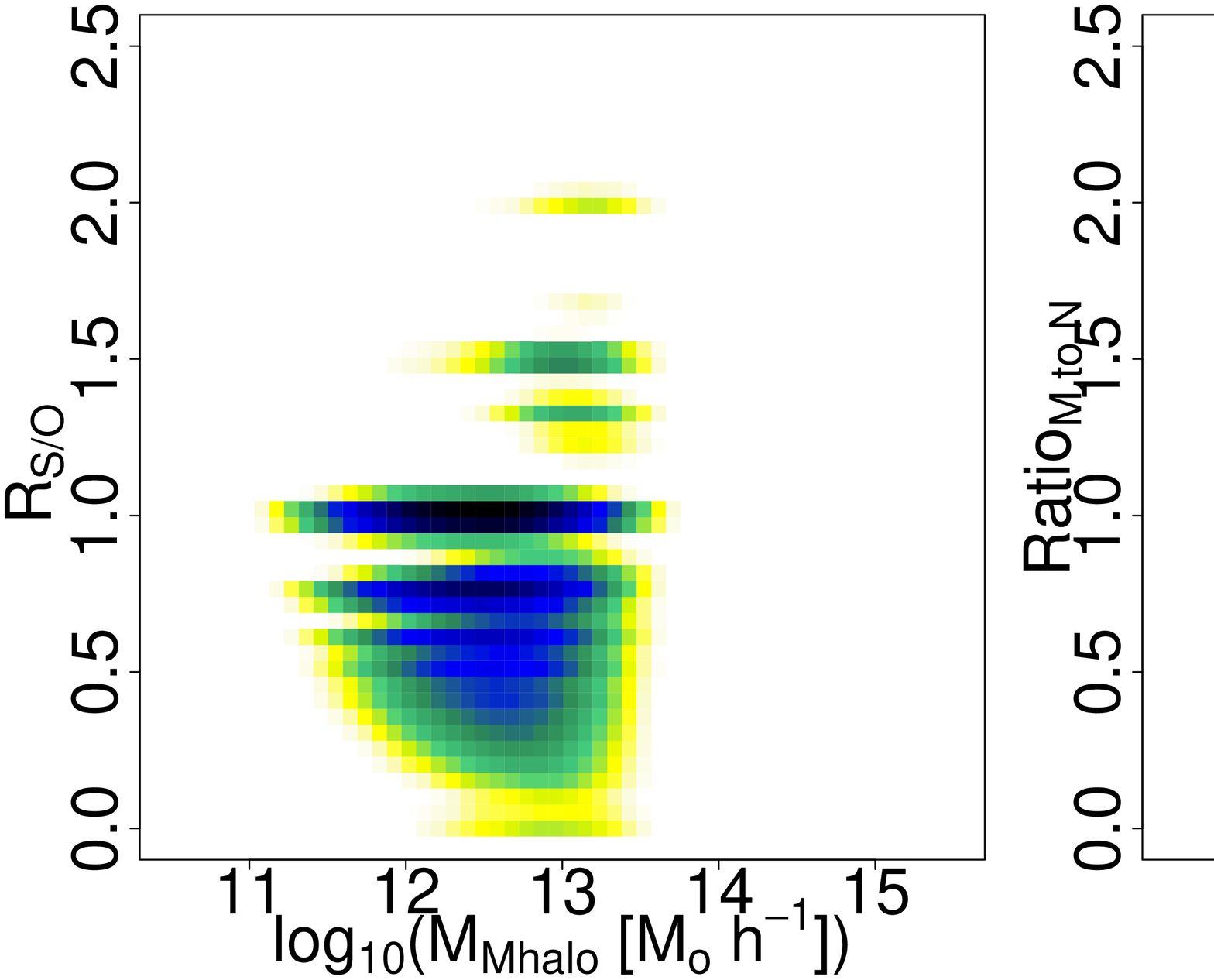}
\includegraphics[clip, width=53.75mm]{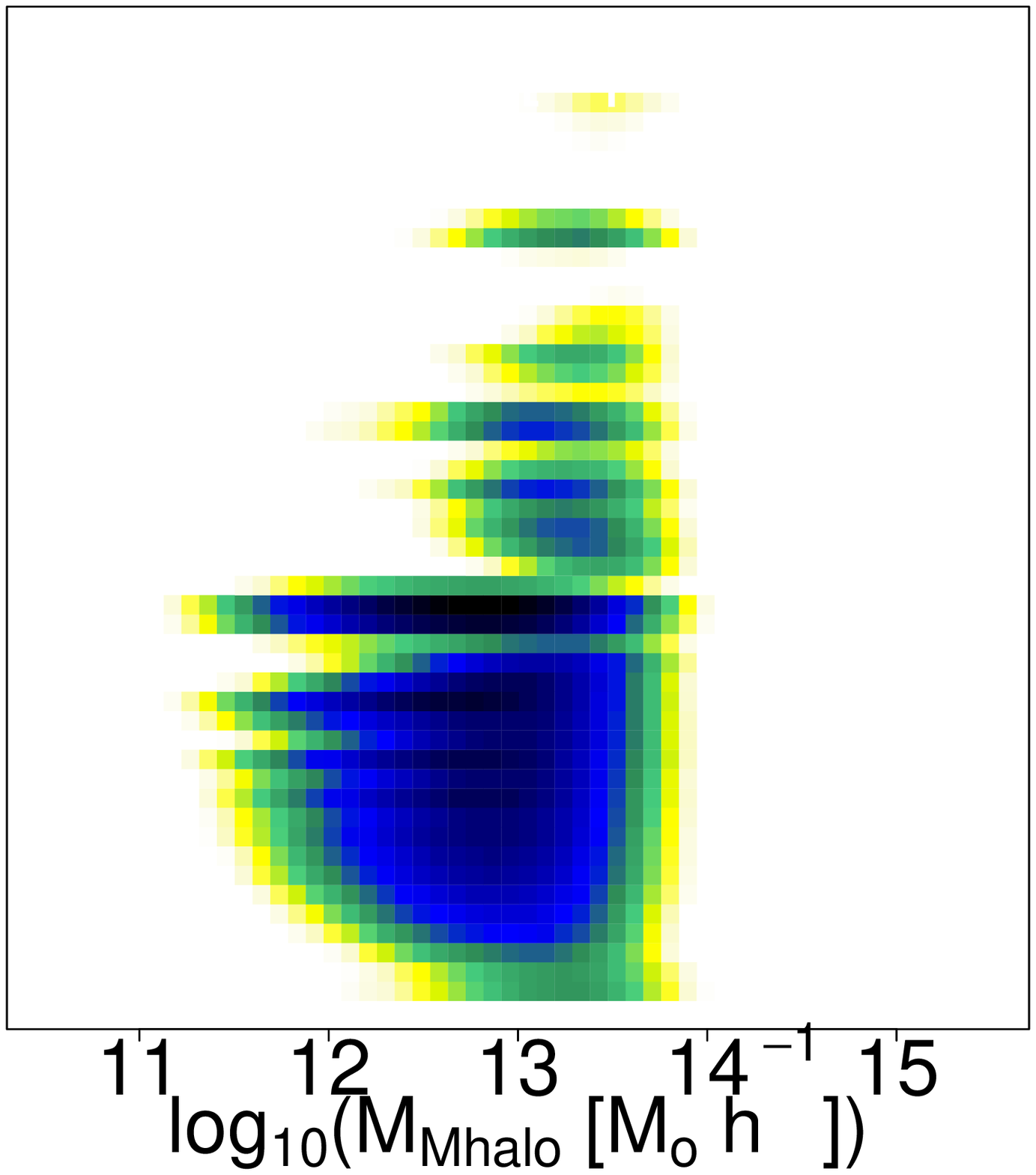}
\includegraphics[clip, width=53.75mm]{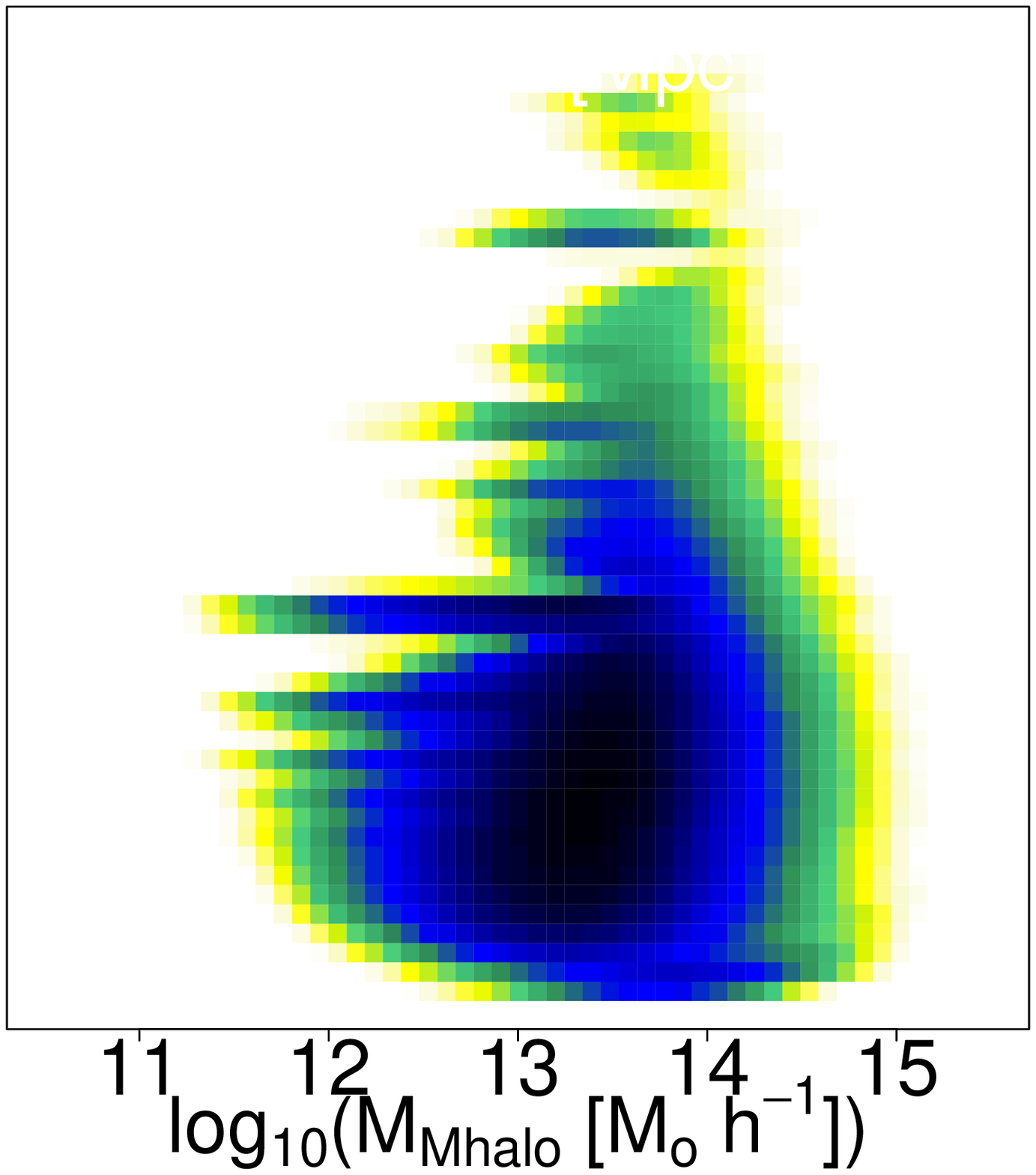}\\
\caption{Smoothed distribution of the ratio between the number of 
satellites (S) and the number halos associated by the method from 
the observational constraints (O), $R_{S/O}$, for these main halos 
as a function of the virial mass in case (A). The three panels 
correspond to equally populated ranges of environmental density,
increasing from left to right. The colour bar varies from yellow 
to black in the same manner as Figure\,\ref{simul}, increasing 
for more probable values.}    
\label{frac211}    
\end{figure*}

Going forward with this analysis, we focus in the fate of halos 
wrongly classified by the method. Since the halo population is
dominated by isolated low mass halos in the field or grouped in pairs
(representing $\approx 93\%$ and $\approx 5\%$ of the main halos,
respectively), we studied separately the sample of main halos 
presenting at least two satellites, assuming this as a simplistic 
threshold to define galaxy groups. 
We define $R_\textrm{S/O}$ as the ratio between 
the number of satellites originally defined by the halo finder over the dark 
matter simulation (S) and the number of halos associated by the 
method from the observable information (O). According to this, 
when $R_\textrm{S/O} > 1$ the method associates too few halos as 
satellites, whereas when $R_\textrm{S/O} < 1$ the method associates 
more satellites than the ones originally defined. 
Using this definition, Figure~\ref{frac211} shows the 
smoothed distribution of $R_\textrm{S/O}$ for the main halos
having more than one satellite,
as a function of the virial mass in case (A). 
The three panels correspond to density ranges with comparable 
number of halos, increasing from left to right.
The colour bar ranges from yellow to black in the same manner than
Figure~\ref{simul}, increasing for more probable values. 

The distribution of ratios in the left panel looks somehow discretised in 
comparison with the other two panels. This is due to the low density of
the environment where these main halos are located, particularly in main
halos with few satellites and less neighbours. The method results successful 
in identifying the members of a larger number of main halos grouped around 
$R_\textrm{S/O}=1$, representing the $40\%$ of the main halos in this 
sample. Around $10\%$ of the main halos were associated to more massive 
ones, and for the $\approx 48\%$ the method associated a larger number of 
halos than their real satellites. There is a small fraction of halos with 
ratios larger than unity, indicating they were identified as main halos but
lost some of their satellites. These halos typically present masses 
larger than $10^{12}\,{\rm M_{\odot}\,h^{-1}}$. On the other hand, the
more massive main halos are able to achieve lower values of 
$R_\textrm{S/O}$. This might be due to the increase in the velocity 
dispersions of their real satellites, favouring the occurrence of fake 
positives. 

The middle panel corresponds to intermediate environmental densities. 
The percentage of main halos with $R_\textrm{S/O} = 1$ 
decreases ($\approx 18\%$) in favour of lower ratios ($\approx 63\%$). 
Also the fractions of main halos stripped of 
satellites or associated to more massive halos increase, representing
the $14$ and $5\%$, respectively.

The right panel corresponds to the highest density environments, typical of rich 
groups and cluster of galaxies. The masses spans a wider range than  
previous panels, and the results point to a lack of accuracy in halos
association. The behaviour of this sample
continues the general trend of the previous panels, but with 
a marginal fraction of main halos around $R_\textrm{S/O}=1$ 
($\approx 7\%$) and many cases of fake positives ($\approx 62\%$), 
with $\mu_\textrm{R}=0.5$ and $\sigma_\textrm{R}=0.22$ the mean and dispersion for main halos presenting $R_\textrm{S/O}<1$. The fraction 
of main halos stripped of their satellites decreases in favour of the 
halos associated to more massive main halos, representing the
$7$ and $25\%$, respectively.

\begin{figure}    
\includegraphics[clip, width=44.5mm]{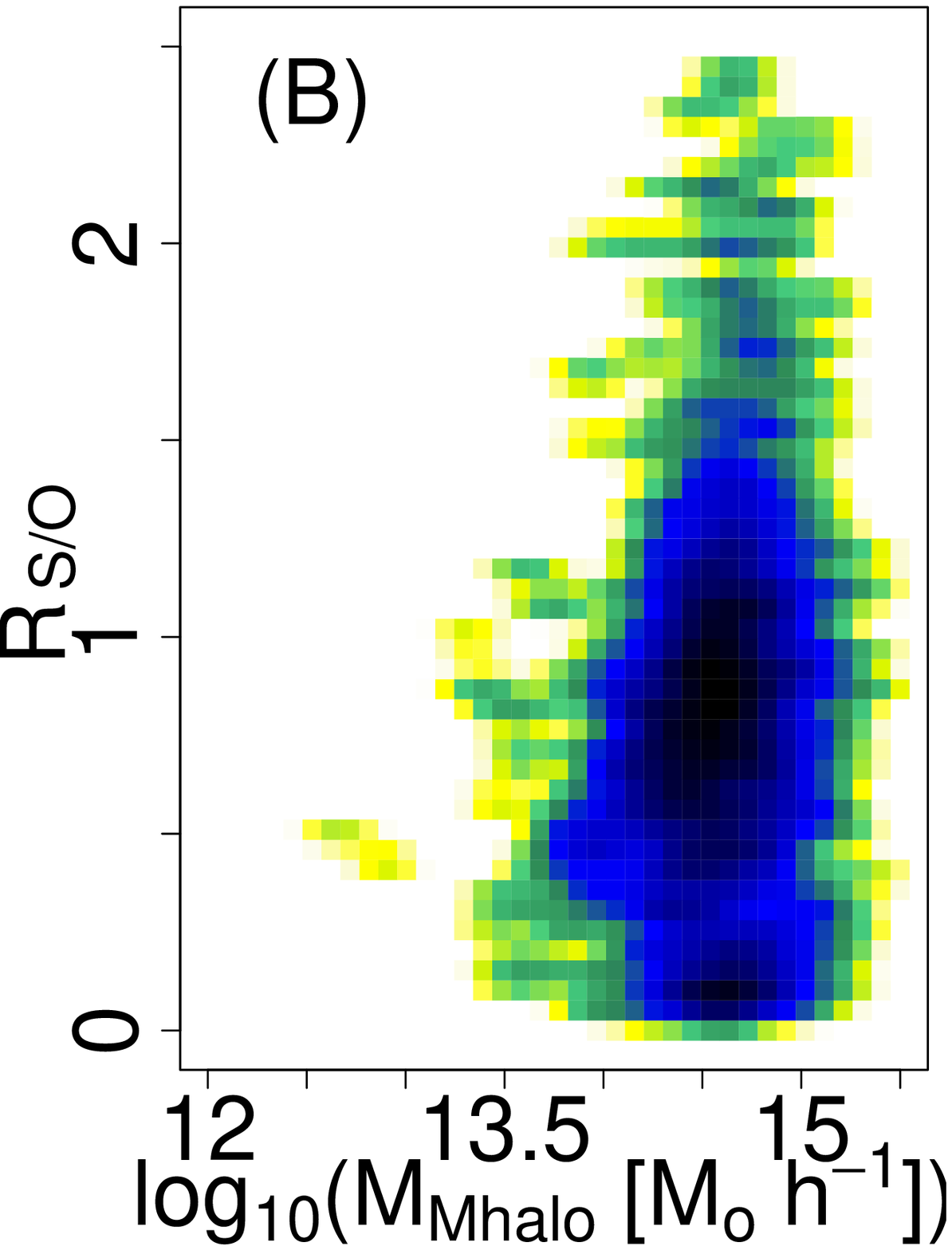}
\includegraphics[clip, width=38.5mm]{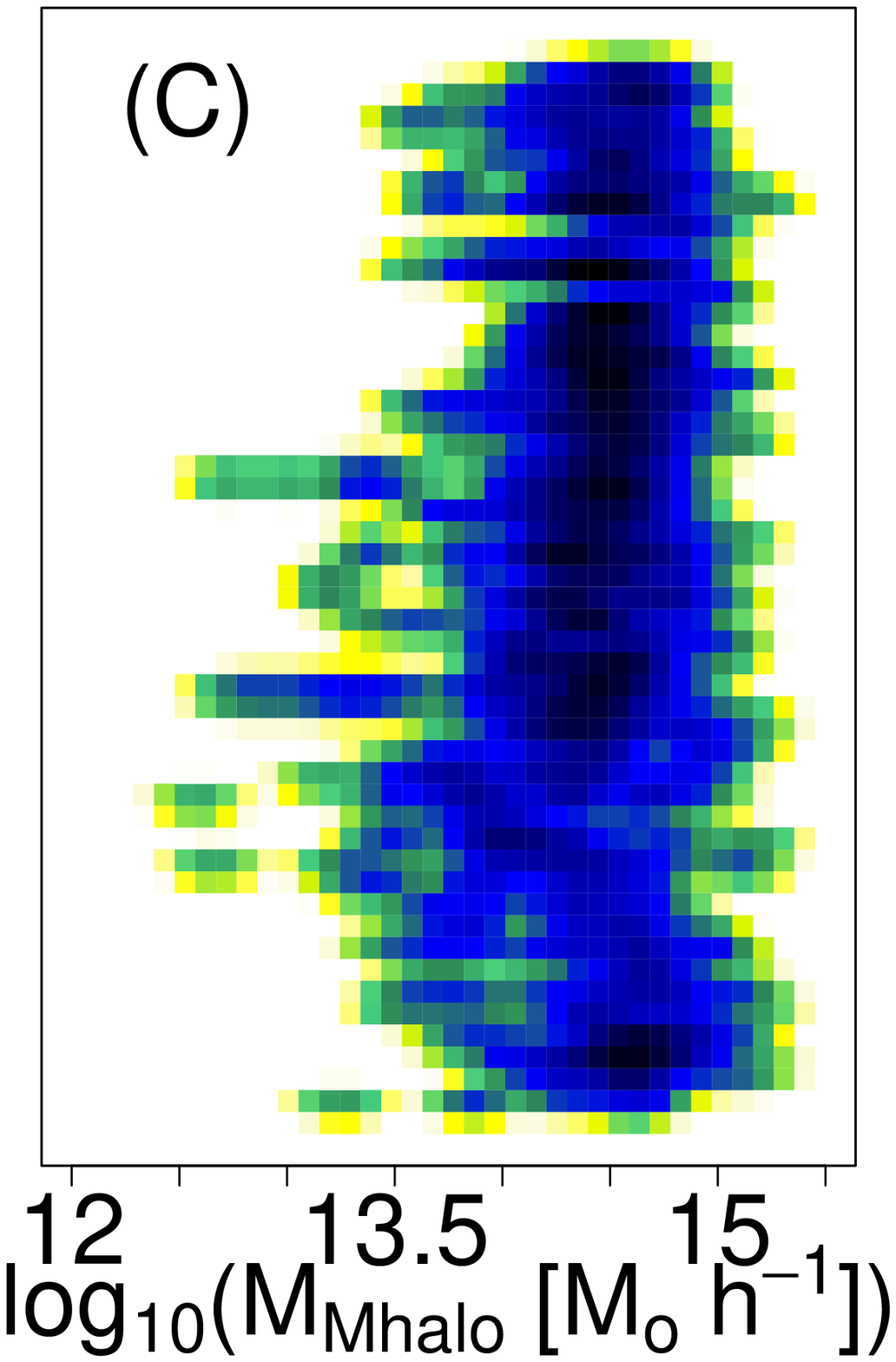}
\caption{Smoothed distribution of the ratio between the number of 
satellites and the number halos associated by the method 
($R_\textrm{S/O}$)
for the densest percentile of halos more massive than 
$10^{12}\,{\rm M_{\odot}\,h^{-1}}$ in cases (B) and (C), left and right
panels, respectively.}    
\label{frac212}    
\end{figure}

Figure~\ref{frac212} is analogue to the panel corresponding to 
the densest percentile in Figure~\ref{frac211}, but for halos 
more massive than $10^{12}\,{\rm M_{\odot}\,h^{-1}}$ in cases (B) 
and (C), left and right panel, respectively. We do not include 
the panels corresponding to sparser environments to limit the 
number of figures. In both cases we found the same differences 
as in Figure~\ref{frac211}.

In case (B) the distribution of main halos resembles the behaviour
from Figure~\ref{frac211} (corresponding to case A). The distribution 
favour discrete values in sparser environments, presenting a large 
fraction of halos around $R_\textrm{S/O}=1$ which drops in denser 
regions from $54$ to $14\%$. On the other hand, main halos with 
$R_\textrm{S/O}<1$ rises from $36\%$ up to $62\%$ in cluster-like 
environments. The percentage of halos striped of satellites or 
that were associated to more
massive main halos also increases with environmental density, 
remaining in all cases below $\approx 12\%$. These latter values are 
lower than those in case (A), indicating that the absence of the 
large number of halos with masses below $10^{12}\,{\rm M_{\odot}\,h^{-1}}$
makes more difficult the occurrence of fake positives.

The behaviour of the ratio differs for case (C) due to the 
different value of $V_\textrm{R,max}$, being 350 instead of 1100\,km\,s$^{-1}$. 
This constrains the percentage 
of main halos with $R_\textrm{S/O}<1$ below $30\%$. Despite of this, 
there is no improvement in the fraction of main halos around $
R_\textrm{S/O}=1$, which ranges from 44 to $12\%$, slightly 
smaller values than in case (B) and similar behaviour towards 
denser environments. The number of main halos accreted by more 
massive ones remains below $5\%$, but there is a large fraction of 
halos with lost satellites (i.e., $R_\textrm{S/O}>1$). This
results are expected since the linking-length parameters chosen for
case (C) favour the correct identification of main halos in detriment
of satellites detection (see lower panel in Figure~\ref{d0v0} and
Section~\ref{sec.param}).

In Figures~\ref{frac211} and \ref{frac212} we focused on main halos
presenting at least two satellites because field halos and those with 
a single satellite dominate the population at low masses. 
Therefore, we now analyse the associations of the halos presenting 
less than two satellites.

\begin{figure}    
\includegraphics[clip, width=\columnwidth]{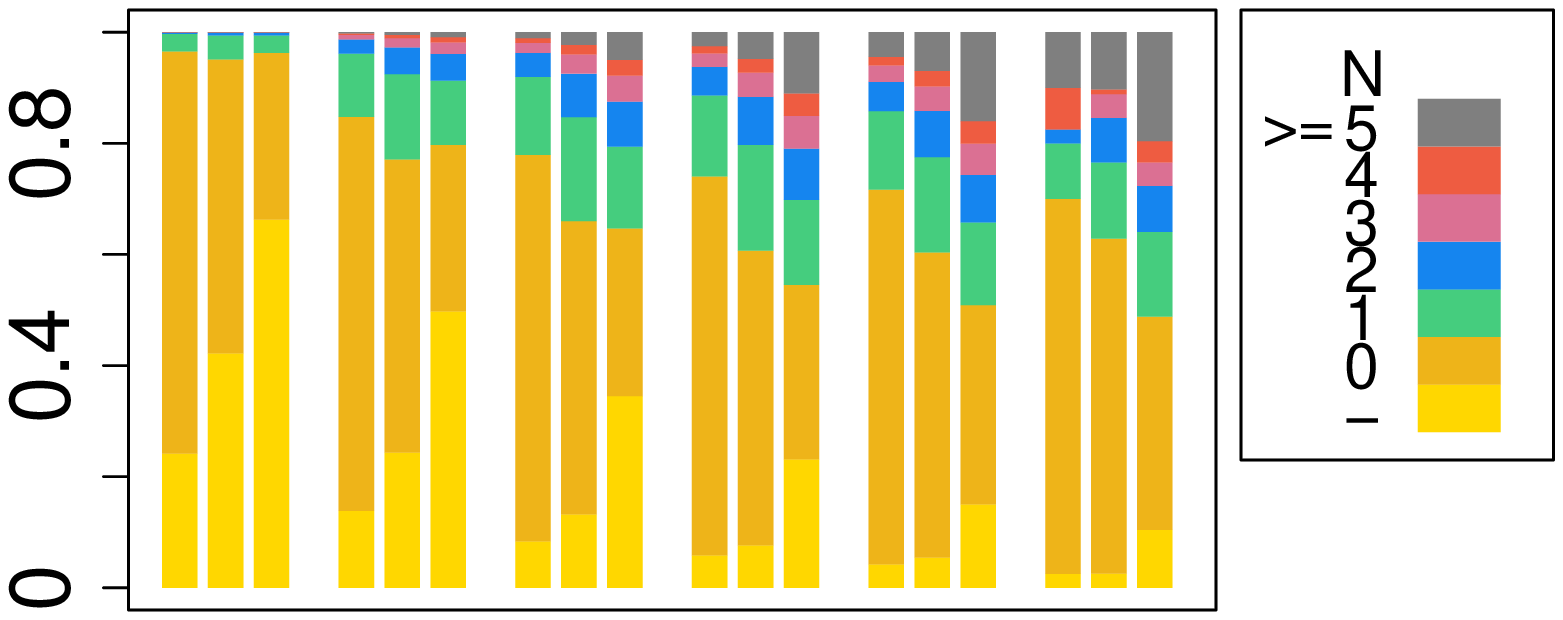}
\includegraphics[clip, width=\columnwidth]{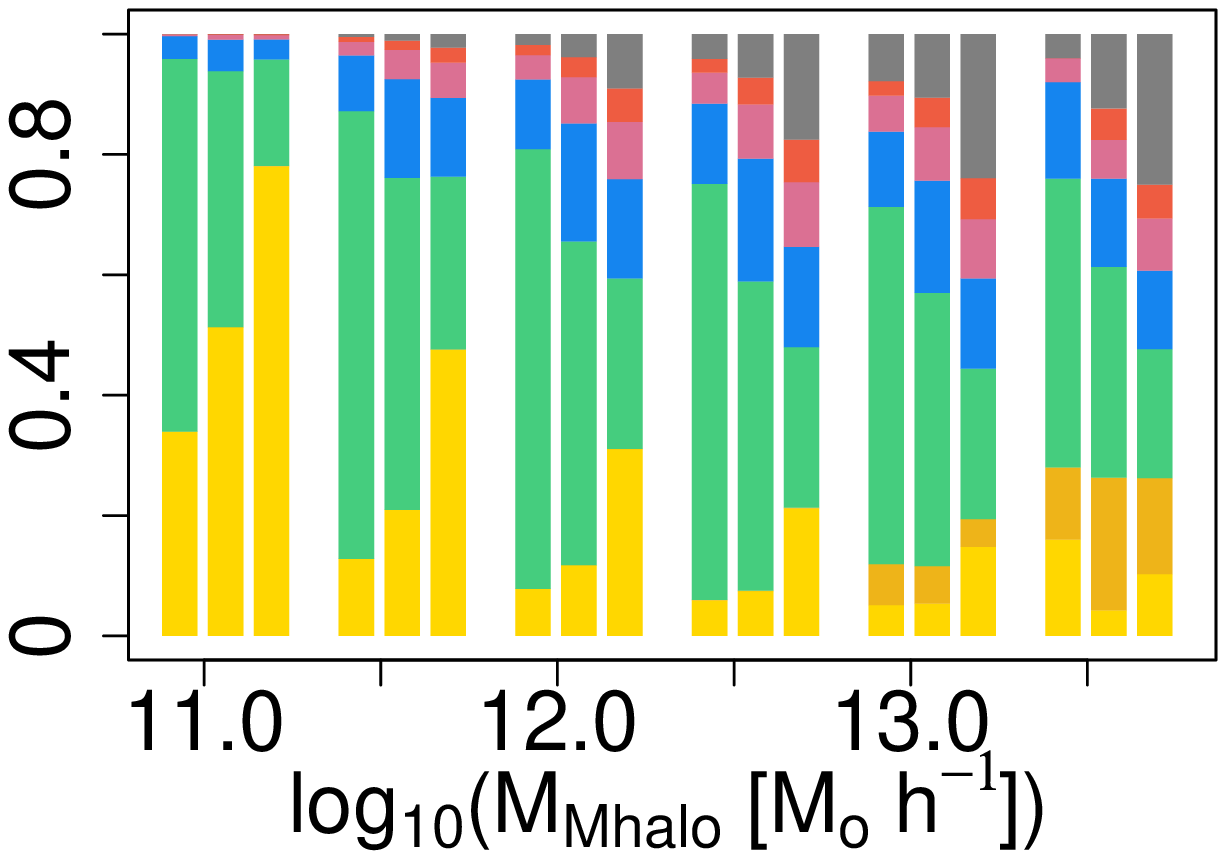}
\caption{Distribution of the number of satellites associated
by the method for field halos (upper panel) and halos with 
a single satellite (lower panel) in case (A). The samples are 
split in six bins of mass and each one in three density bins.
The colour labelled with a dash correspond to halos associated
by the method to more massive ones.}    
\label{frac311}    
\end{figure}

The Figure~\ref{frac311} shows the distribution of the number of 
satellites associated by the method for different main halo masses and 
environmental densities. We split the sample 
in six mass bins and each one in three bins of increasing environmental 
density, with the first one corresponding to isolated galaxies, and the 
third one to the surroundings of a group/cluster of galaxies. 
The cuts to define the ranges of environmental density are chosen 
such that its bins are equally populated. 
The number of associated satellites is depicted with 
different colours, and in each bin they are organised 
in the y--axis according to the
corresponding cumulative fraction with respect to the total. 
The upper panel of the figure corresponds to fields halos in case (A).
Independently of the virial mass, the method is correctly classifying
(those labelled with ``0'') $\approx 70\%$ of the halos located 
in low density environments, with this fraction decreasing towards denser 
environments. Depending on the virial mass, misclassified halos will be 
associated to a more massive one (labelled with a dash in the plot) or
will ``acquire'' fake satellites.
 
The lower panel of the Figure~\ref{frac311} is analogue to the 
upper one but only considering main halos with a single satellite. Once
again, the method is more successful in low density environments, with
slightly better results for intermediate mass halos. Except for
the most massive bin, the number of halos classified as field ones is
negligible. This occur because main halos with masses below few times 
$10^{13}\,{\rm M_{\odot}\,h^{-1}}$ present typical separations with their 
satellites in both projected distance and radial velocity smaller than
$D_\textrm{p,max}$ and $V_\textrm{R,max}$, 
respectively (see middle-right and right panels
in Figure\,\ref{simul}). Hence, the chosen linking-length parameters 
restrict the association outcome of main halos and satellites. As expected, the most critical scenario takes place for
halos in dense environments, with only $\approx 
20\%$ of the cases being correctly classified.

The results when the analysed sample is restricted to halos more massive than 
$10^{12}\,{\rm M_{\odot}}$ varies depending on the linking-length
parameters. The absence of halos in the mass range
$10^{11} - 10^{12}\,{\rm M_{\odot}}$ leads to larger mean distances 
for halos, driving to better results for larger $D_\textrm{p,max}$ 
(lower panel Figure~\ref{d0v0}). In case (B) the distribution presents a 
similar behaviour as in Figure\,\ref{frac311}, with minor changes in the 
percentages, for both field halos and those with a single satellite.
In case (C) the fraction of properly classified field halos is larger
and the typical number of fake satellites associated to massive halos 
is smaller. It also decreases the fraction of low mass main halos
wrongly associated to more massive halos. For main halos presenting
a single satellite results are also different in cases (B) and (C).
In this latter case the method is successful for a larger fraction of 
main halos when their virial masses are
lower than $10^{13}\,{\rm M_{\odot}\,h^{-1}}$,
but it classifies as field halos a considerable fraction of halos
above this limit. This is probably due to the restrictive value of 
$V_\textrm{R,max}$ considered in case (C). 

\subsection{Comparing observational properties of real satellites and associated halos}

\begin{figure}    
\includegraphics[clip, width=\columnwidth]{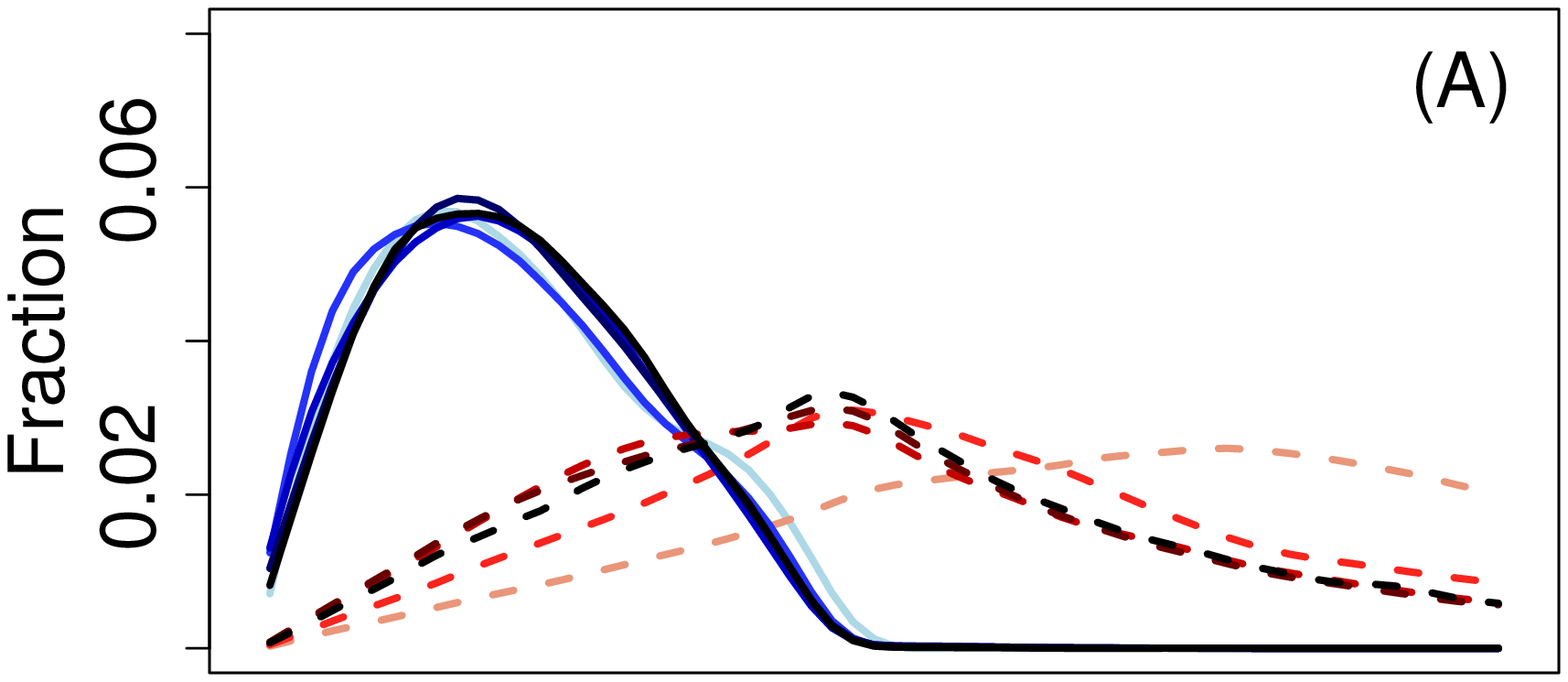}\\
\includegraphics[clip, width=\columnwidth]{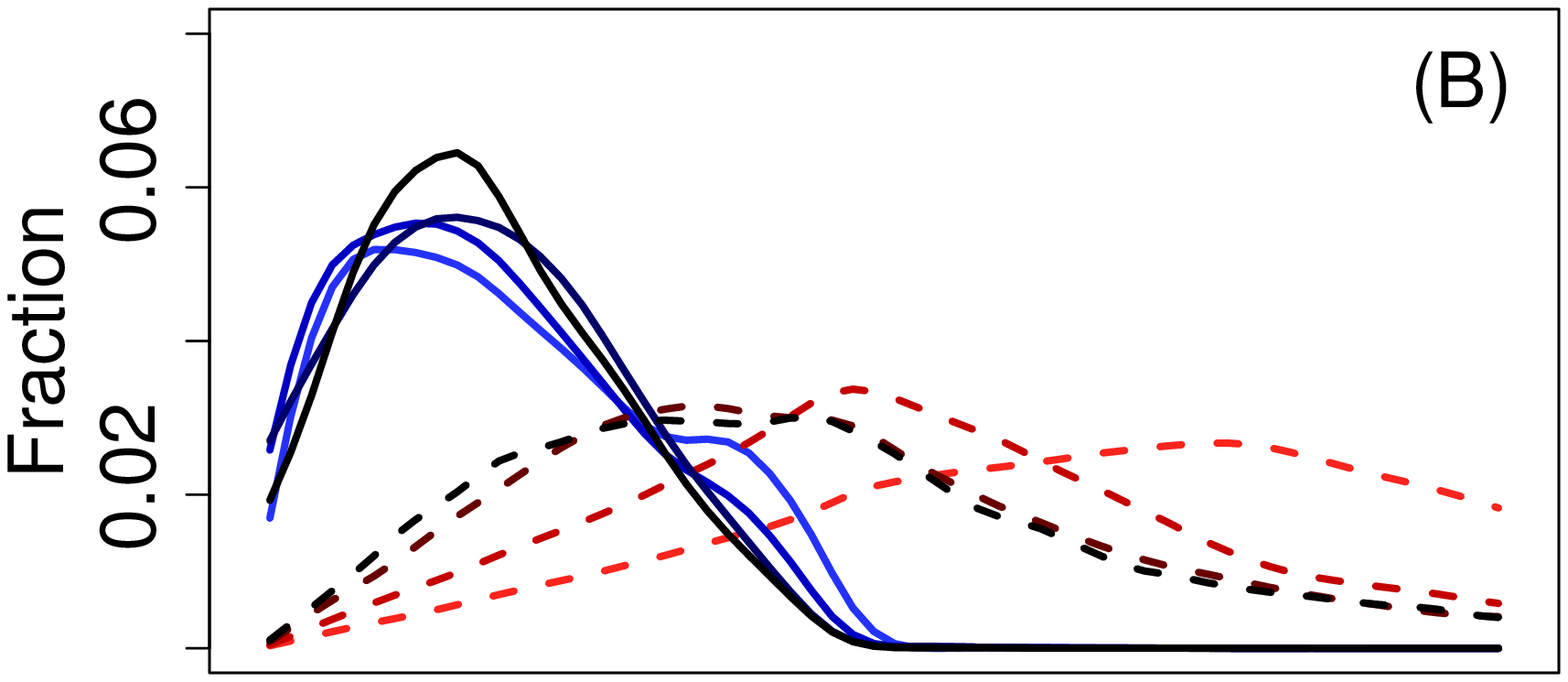}\\
\includegraphics[clip, width=\columnwidth]{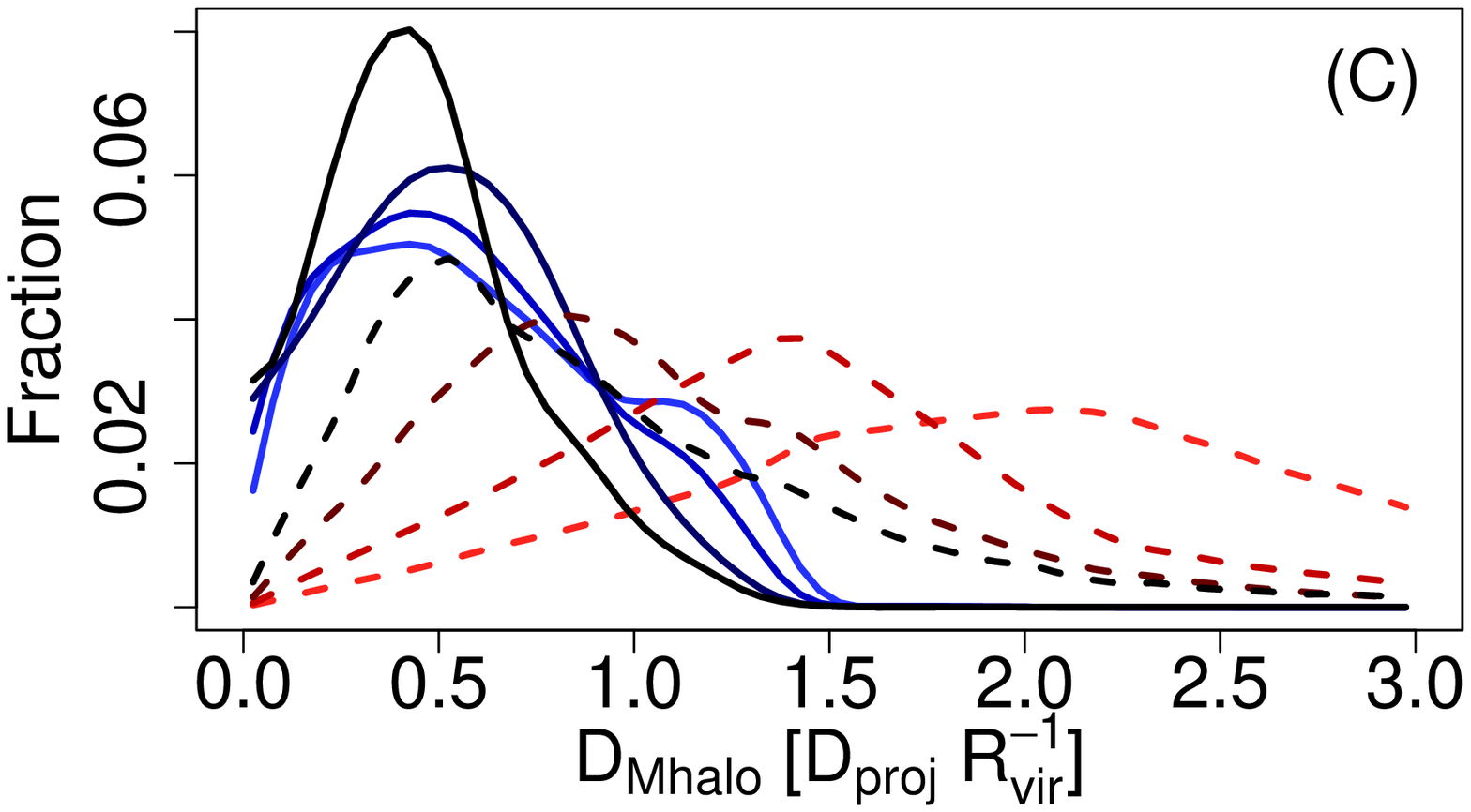}\\
\caption{Distribution of projected distances to the main halo associated 
by the method for real satellites (solid lines) and fake positives (dashed
lines). The gradient in colour scheme corresponds to mass intervals for the 
main halos, from the least (light) to the most massive ones (dark). The upper
corresponds to the case of halos more massive than $10^{11}\,{\rm M_{\odot}}$,
while middle and lower panels correspond to the two case analyzed for
halos more massive than $10^{12}\,{\rm M_{\odot}}$.}    
\label{proj}    
\end{figure}

Henceforth we analyse the impact on the derived observational 
properties of galaxies due to the method performance for the cases 
selected in Section~\ref{sec.param}. As a first approach, 
Figure\,\ref{proj} shows the distribution of projected distances
to the main halo associated by the method for real satellites 
(solid lines) and fake positives (dashed lines). The gradient in 
colour corresponds to mass intervals for the main halos, from the 
least (lighter) to the most massive ones (darker).
The projected distances are normalised by the virial radii, to facilitate
the comparison between different mass intervals. In the three cases,
most of
the satellite distributions reach their upper limit at 
$R_{\rm proj} \approx 1.5 R_{\rm vir}$, which is in agreement with 
the results from Section\,\ref{simstats} and Figure\,\ref{simul}.

The upper panel corresponds to results when the method is applied to 
halos more massive than $10^{11}\,{\rm M_{\odot}}$ (case A). In this case,
the distributions of
fake positives grow with the projected distance up to
$R_{\rm proj} \approx 1.5 R_{\rm vir}$, where distributions start to decline
due the lack of ``real'' satellites to be associated with. The only 
exception is the distribution of the less massive main halos, which 
grows up to $\approx 3 R_{\rm vir}$. This occurs because the mass interval 
spans virial masses in the range $\approx 10^{11} - 10^{12}\,{\rm M_{\odot}\,
h^{-1}}$, whose typical virial radii are much lower than the linking-length
in projected distance $D_{\rm p,max}= 525$\,Mpc, i.e., $\approx 355\,{\rm Mpc\, 
h^{-1}}$.

A similar behaviour can be found in the middle panel, which corresponds to 
the case (B). In this case the linking-length is $\approx 660\,
{\rm Mpc\,h^{-1}}$, comparable with the virial radius for a main halo of
$\approx 5 \times 10^{13}\,{\rm M_{\odot}\,h^{-1}}$. 
Likewise, the lower panel shows the results from case (C).
The observed differences with the previous ones are a consequence of
the more restrictive linking-length in radial velocities ($V_{\rm R,max}$)
used in this case. For the most massive main halos, this choice translates
in a lost of satellites in the outskirts (dark solid curve), resulting
in a narrow distribution in comparison with case (B). This was already 
noticed in Figures\,\ref{frac112} and \ref{frac212}. The choice of $V_{\rm R,max}$
also modified the distributions of fake positives,   
shifting the location of their maximum
towards smaller projected distances (in terms of the virial radii).

\begin{figure}    
\includegraphics[clip, width=\columnwidth]{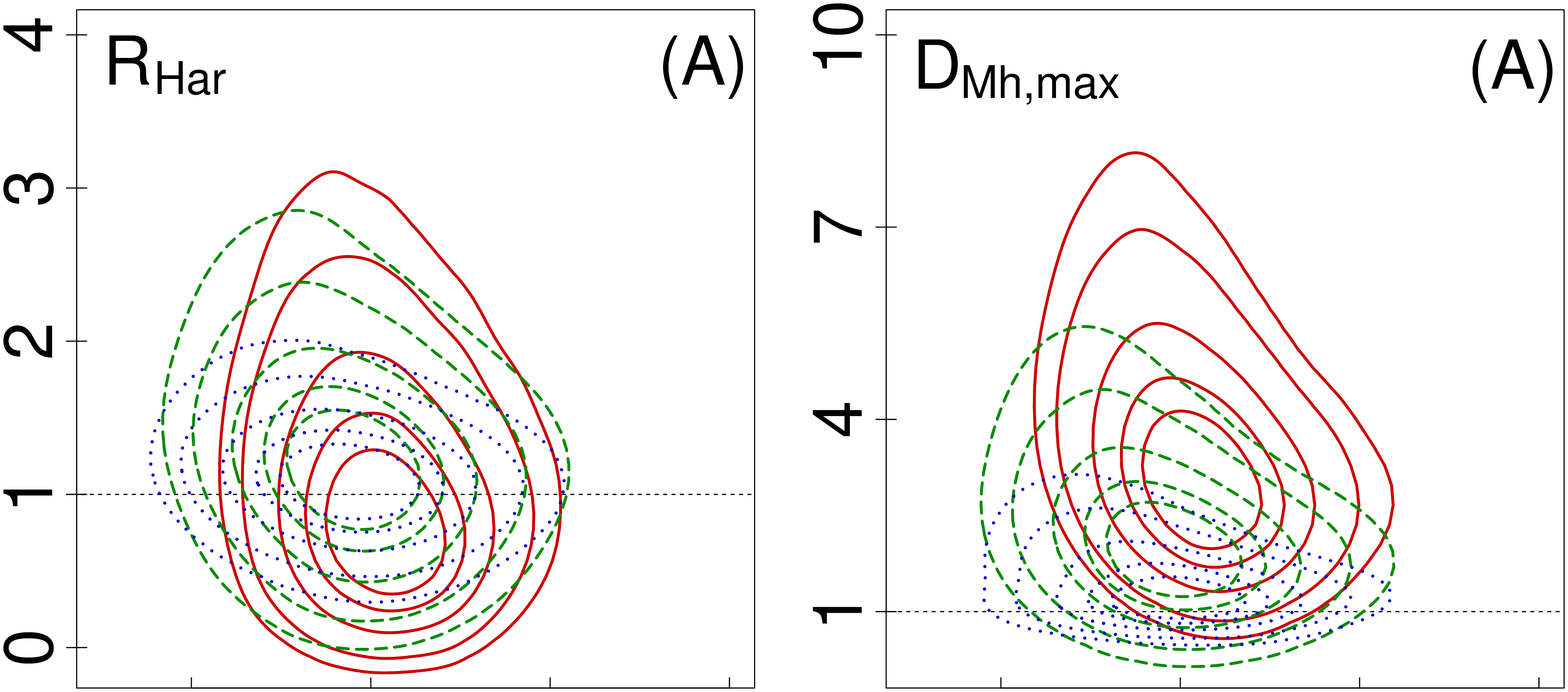}\\
\includegraphics[clip, width=\columnwidth]{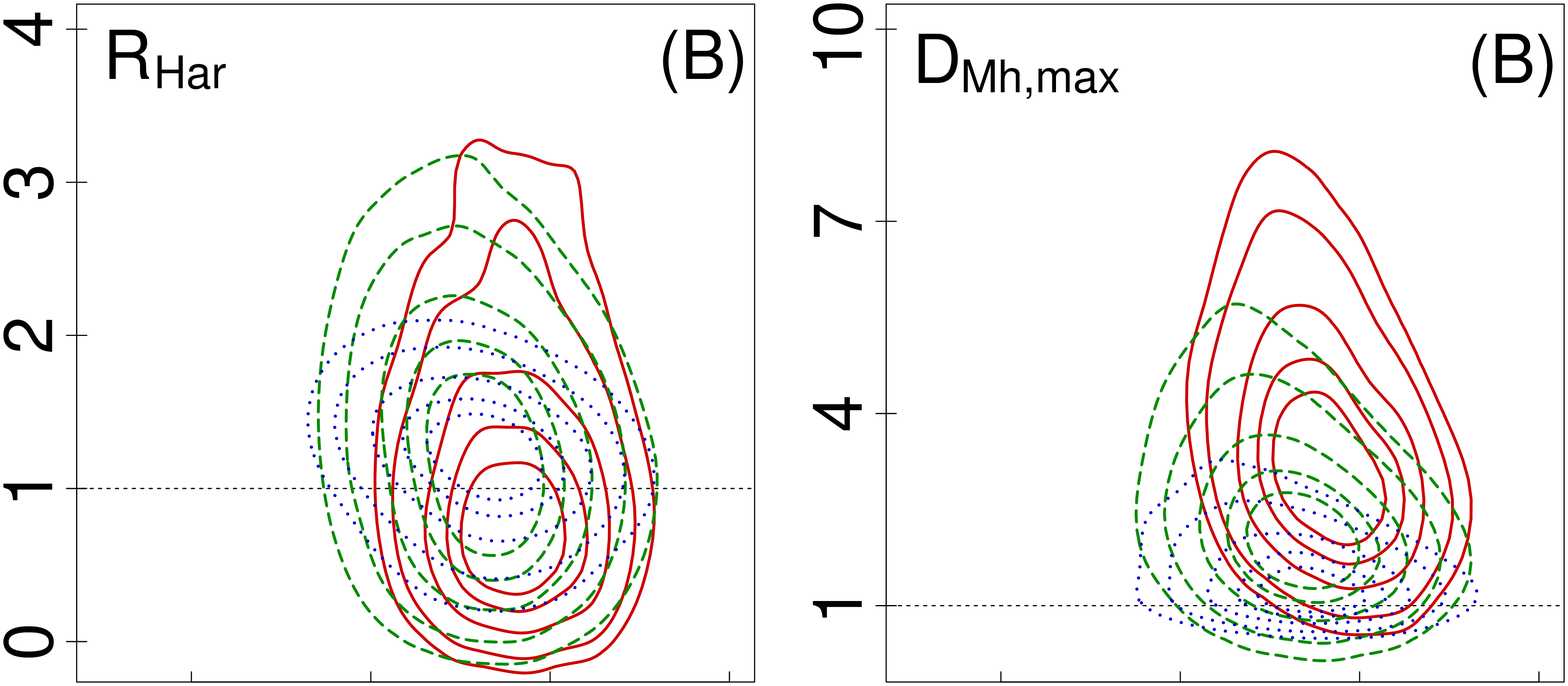}\\
\includegraphics[clip, width=\columnwidth]{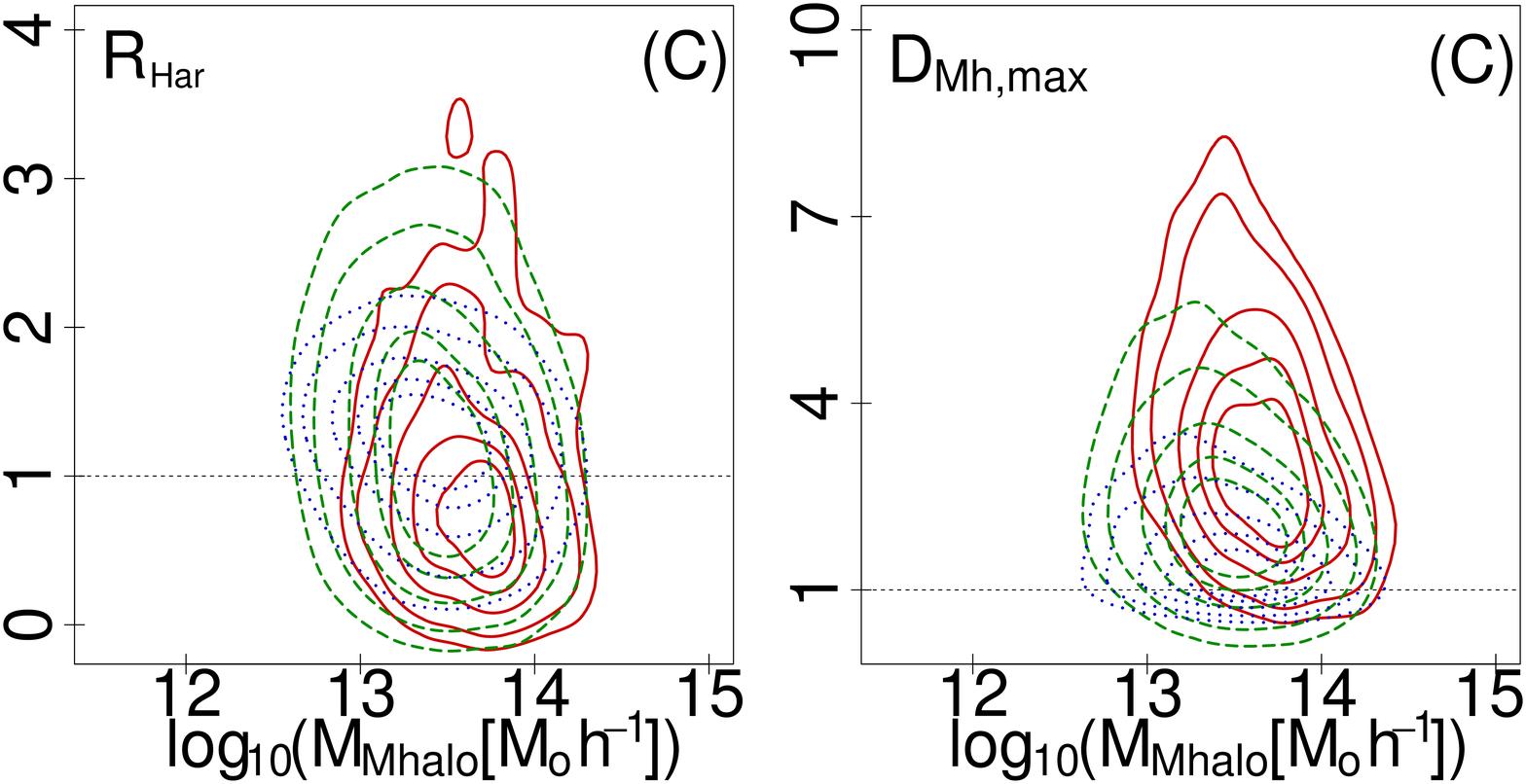}\\
\caption{{\bf Left panels:} Ratio between the harmonic radius 
associated by the method for galaxy groups, and their
real harmonic radius, for the three cases analysed in this paper. 
Solid contour levels correspond to main halos
with $R_{\rm S/O}<0.33$, dashed ones to $0.33<R_{\rm S/O}<0.66$, and
dotted curves represent those with $0.66<R_{\rm S/O}<1$.
{\bf Right panels:} Ratio between the projected distances of 
the farthest halo associated by the method to that of the farthest real 
member, applying the same line code as previously indicated.}
\label{proj2}    
\end{figure}

The harmonic radius in groups of galaxies is associated with the
spread in projected spatial distance between the members. The left 
panels in Figure\,\ref{proj2} shows the ratio between the harmonic 
radius of galaxy groups associated by the method and the real 
harmonic radius of the groups, with this property as defined by 
\citet{fir06}. As previously indicated, we consider the main halos 
with at least two satellite halos as a simplistic definition of 
galaxy groups. The sample is split in three bins of $R_{\rm S/O}$, 
presenting main halos largely contaminated ($R_{\rm S/O}<0.33$) 
with solid contour levels, moderately contaminated ($0.33<R_{\rm S/O}<0.66$) 
ones with dashed curves, and less contaminated main halos 
($0.66<R_{\rm S/O}<1$) with dotted contour levels. The rows 
correspond to the three cases analysed in this paper. Those main 
halos largely contaminated show a wider distribution of ratios 
in cases (A) and (B), with the dotted distribution more 
concentrated towards ratio $\approx 1$. In case (C) the ratio 
of the harmonic radius for largely contaminated halos extend 
up to smaller values in comparison with case (B), due to the 
its more restrictive value of the linking-length $V_{\rm R,max}$, 
which is particularly important for massive main halos (see 
Figure\,\ref{frac212}). The right panels correspond to the ratio 
between the projected distances of the farthest halo associated
by the method to that of the farthest real member, separated in 
bins of $R_{\rm S/O}$ in the same manner than in the other panels.
This parameter exhibits a larger dispersion than the harmonic 
radius in all the cases, particularly in the cases of main halos
largely contaminated.

\begin{figure}    
\includegraphics[clip, width=\columnwidth]{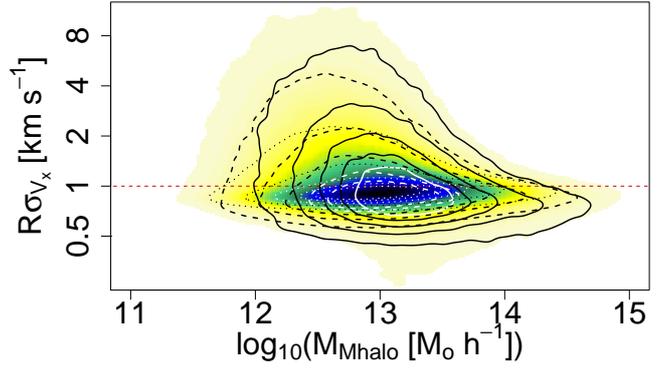}\\
\caption{ Distribution of the ratio between the 
velocity dispersion of main halos (R$\sigma_{\rm V_x}$) derived from halos 
associated by the method and 
the obtained from the
real satellite halos, for the case (A),
depicted with filled coloured contours.
A value larger than 1 indicates the dispersion
obtained from this latter halos is larger than the original.
We also split
the sample in three bins of $R_{\rm S/O}$ represented with different contour
curves in the plot, considering main halos largely contaminated ($R_{\rm S/O}
<0.33$, solid curves), moderately contaminated ($0.33<R_{\rm S/O}<0.66$, 
dashed curves) and less contaminated ($0.66<R_{\rm S/O}<1$, dotted curves).}    
\label{disp1}    
\end{figure}

The distribution of radial velocities is useful to determine the probability
of belonging to a group/cluster for galaxies with close projected distances.
Nonetheless,
the occurrence of fake positives in the association method can
produce changes in the velocity dispersions observationally obtained for 
a main halo (i.e., ``group/cluster of galaxies'') from their assumed members.
Figure~\ref{disp1} shows the ratio between the velocity dispersion derived
for halos associated by the method and their real velocity dispersion
(R$\sigma_{\rm V_x}$) when halos more massive than $10^{11}\,{\rm M_{\odot}}$ 
are considered (i.e., case A). A value larger than one indicates the 
dispersion increases when we considered fake positives from the method.
Its general distribution
is depicted using filled coloured contours, where the 
darker ones represents the most populated levels. Moreover,
we split the sample in 
three bins of $R_{\rm S/O}$ represented with different contour curves in the 
plot, considering main halos largely contaminated ($R_{\rm S/O}<0.33$, solid 
curves), moderately contaminated ($0.33<R_{\rm S/O}<0.66$, dashed curves) and 
less contaminated ($0.66<R_{\rm S/O}<1$, dotted curves). As expected, main 
halos largely contaminated show a wide and asymmetric distribution of ratios.
The fraction of main halos with R$\sigma_{\rm V_x} >1$ decreases towards 
massive halos, reaching 0.5 at $\approx 10^{13}\,{\rm M_{\odot}\,h^{-1}}$. 
This mass roughly correspond to $\sigma_{\rm V_x} \approx 330$\,km\,s$^{-1}$. 
Less massive main halos exhibit smaller dispersions, and the assumed linking-length 
$V_{\rm R,max}= 980$\,km\,s$^{-1}$ lead to an increase in the velocity 
dispersions due to fake associations of satellites. On the other hand, the 
majority of main halos at the massive end present R$\sigma_{\rm V_x} <1$.

\begin{figure}    
\includegraphics[clip, width=\columnwidth]{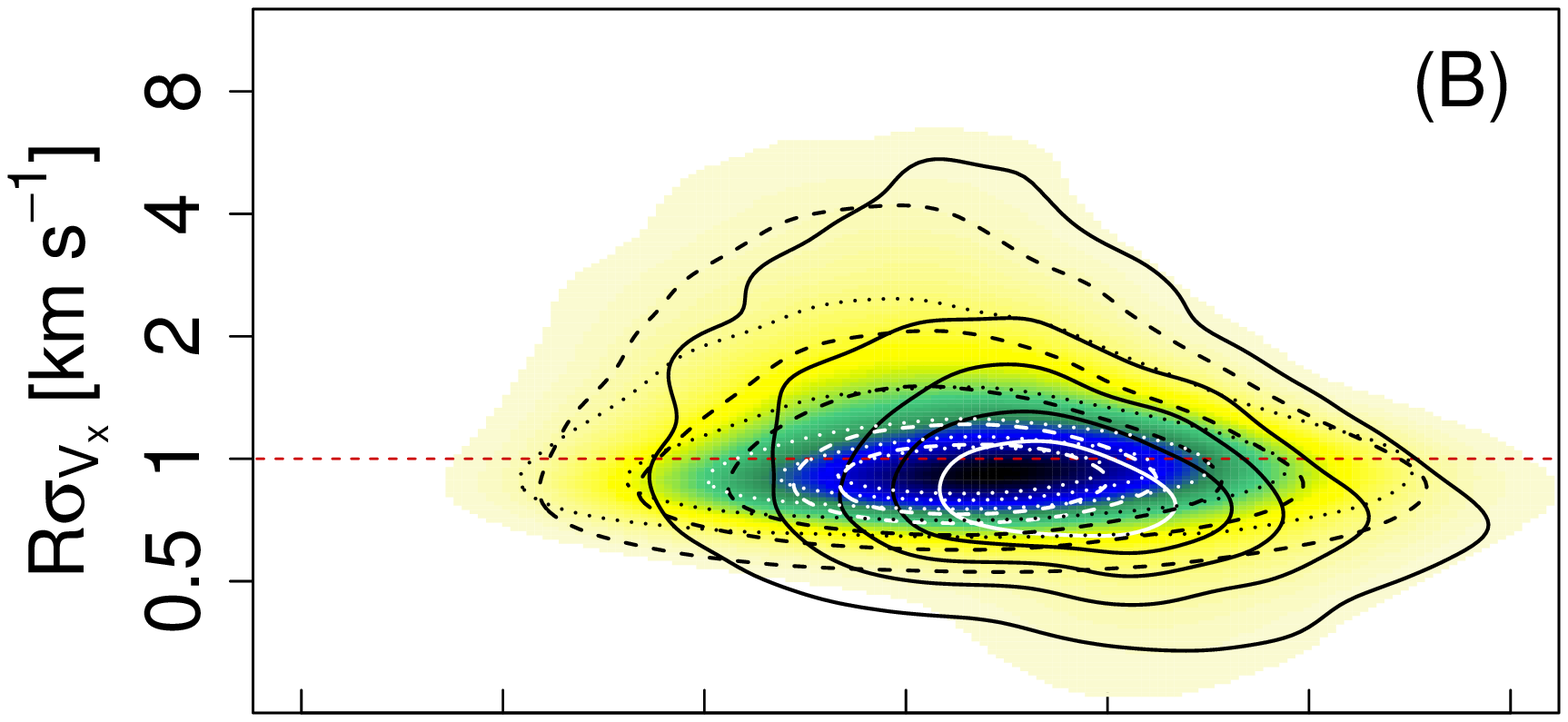}\\
\includegraphics[clip, width=\columnwidth]{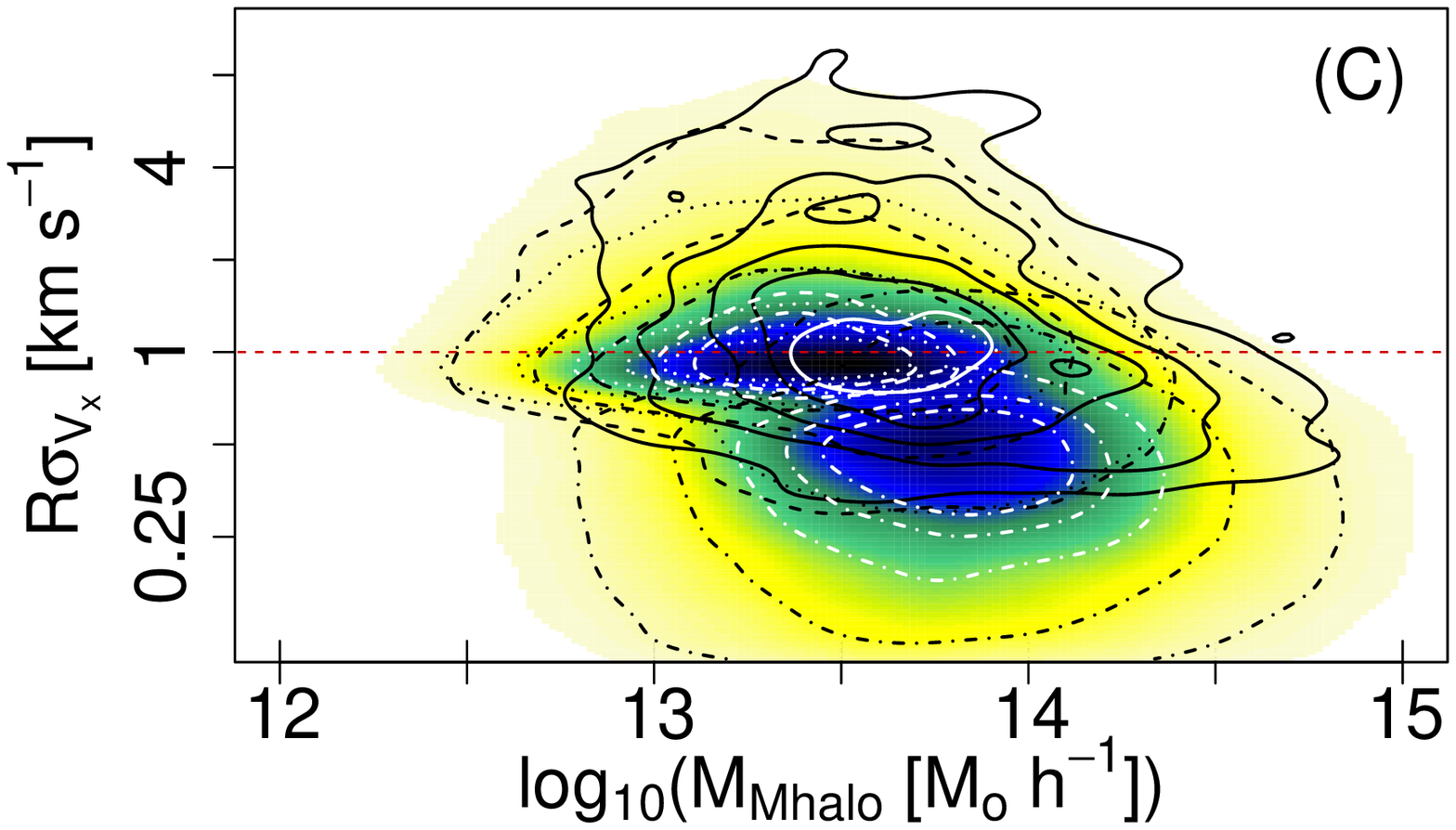}\\
\caption{Distribution of velocity dispersion ratios, analogue to Figure\,\ref{disp1} but for halos more massive than $10^{12}\,{\rm M_{\odot}}$ separating the cases (B) and (C).
For this latter case, the main halos with $R_{\rm S/O}>1$ are also indicated 
in the diagram with dash--dotted contour curves.}    
\label{disp2}    
\end{figure}

Similar distributions are shown in Figure~\,\ref{disp2} 
for both linking-length 
choices when halos more massive than $10^{12}\,{\rm M_{\odot}}$ are considered. 
The behaviour is clearly different in cases (B) and (C), with this latter one 
presenting a bulk of main halos with R$\sigma_{\rm V_x} <1$ spread over a 
large range of masses. The sample was split in three bins in the same manner as
Figure\,\ref{disp1}, and the contour curves follow the same prescription. 
Considering main halos ranging $0<R_{\rm S/O}<1$ do not describe the 
complete distribution in case (C), we added a fourth bin corresponding to
$R_{\rm S/O}>1$ 
with dash--dotted lines. This is in agreement with the right panel of 
Figure\,\ref{frac212}, where a large number of main halos were located in 
this range, and it is responsible for the existence of this second bulk. 
The narrow linking-length assumed in case (C) lead to strip many satellites 
with large peculiar velocities from their
main halos, with the subsequent decrease in the ``observed'' 
velocity dispersion.

The luminosity function is a well studied property in 
groups/clusters of galaxies 
\citep[e.g.][and references there in]{sch76,ric18},
having relevance as a constraint in numerical simulations of galaxy evolution
\citep[e.g.][]{gon14,cor18}. 
In particular, the conditional luminosity functions 
of satellites and central galaxies 
are more affected by the association method, as halos incorrectly 
classified can highly contribute to the number counts of galaxies
for main halos in specific virial mass ranges.
The $K$ magnitude allocation described
in Section~\ref{mag.alloc} order the halos in terms of their virial 
masses without discrimination between field or satellite ones. Because
galaxy evolution might depend on environmental factors, our luminosities 
are only indicative, and we do not pretend to 
directly compare results from this Section with observational neither 
numerical ones. However, the comparison between luminosity functions 
from real cluster members and halos associated by the method in this 
analysis is relevant to understand how dependent are observational 
results from the association of galaxies.

\begin{figure}    
\includegraphics[clip, width=\columnwidth]{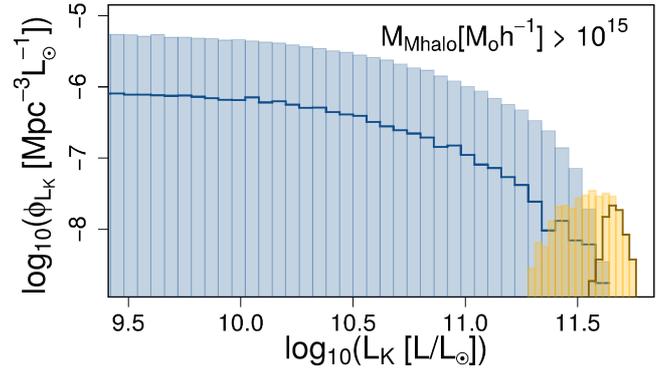}\\
\caption{
Conditional luminosity functions of galaxies belonging to 
main halos with virial mass larger than $10^{15}\,{\rm M_{\odot}\,h^{-1}}$.
Systems obtained from the percolation method 
in case (A) are shown in filled histograms, whereas the 
expected luminosity functions corresponding to the systems defined 
by the simulation are shown in solid thick lines histograms.
Satellites and centrals are represented separately 
in blue and yellow histograms, respectively. 
}
\label{fl11}    
\end{figure}

Figure\,\ref{fl11} shows conditional luminosity functions for
clusters of galaxies. These are obtained by selecting the 
main halos with virial masses larger than
$10^{15}\,{\rm M_{\odot}\,h^{-1}}$ when the systems 
associations with the percolation method in the case (A)
are considered (filled histograms), 
and they are compared with the clusters defined by
the original simulation (solid thick line histograms).
In both cases, the central galaxies (i.e, the luminosities
assigned to the main halos) and satellites are presented
separately with yellow and blue histograms, respectively.
The luminosity function of central galaxies
is significantly different in both cases, with a large
contribution to lower luminosities of main halos achieving virial masses
above $10^{15}\,{\rm M_{\odot}\,h^{-1}}$ when the method is applied. 
This behaviour is a consequence of 
the large number of fake satellites that are assigned by the method
(these correspond typically to main halos with $R_{\rm S/O}<1$ in the right 
panel of Figure\,\ref{frac211}). Therefore, when results from the percolation
method are assumed, the number of systems in this mass range is $\approx 4$ 
times larger than expected. 
These additional main halos constitute the population of galaxies that fills
the fainter part of the distribution of luminosities.
On the other hand,
the number of satellites associated to these systems by the method is
$\approx 7$ times larger, which implies that the number of satellite 
halos associated is $\approx 1.8$ times the mean number of satellites
for main halos in this mass range. This overestimation of satellites 
contributes to the luminosity function homogeneously, since no difference
is expected in the shape of the luminosity function for satellites with
respect to the main halos virial mass.

We repeated this analysis for decreasing virial mass ranges 
$10^{14} < M_{\rm vir} < 10^{15}\,{\rm M_{\odot}\,h^{-1}}$ and 
$10^{13} < M_{\rm vir} < 10^{14}\,{\rm M_{\odot}\,h^{-1}}$.
The number of main halos in these mass ranges from the percolation method
in terms of the real number of main halos decreases to $\approx 1.5$ and
$\approx 1$, respectively. In both cases there is a paucity of main halos
in the bright end of the distribution, and an excess of faint ones, both
resulting from the contribution of fake satellites to the system virial 
mass. In these cases there also exist an overpopulation of 
satellites, which contributes homogeneously over the entire distributions,
resulting $\approx 2.7$ and $\approx 2$ times the number of the originally 
bounded satellites, respectively. Hence, the mean excess of satellites
per main halo are $\approx 1.8$ and $\approx 2$, respectively. This 
indicates an approximately steady mean fraction of fake positives for 
main halos more massive than $10^{13}\,{\rm M_{\odot}\,h^{-1}}$, which 
is in agreement with the distribution shown in the right panel of
Figure\,\ref{frac211}.

On the other hand, results when halos more massive than $10^{12}\,{\rm M_{\odot}}$ 
are considered differ depending on the linking-length parameters (cases B and C). 
In case (B) we obtained a similar behaviour than previously indicated for case (A)
for the three mass bins.

\begin{figure}    
\includegraphics[clip, width=\columnwidth]{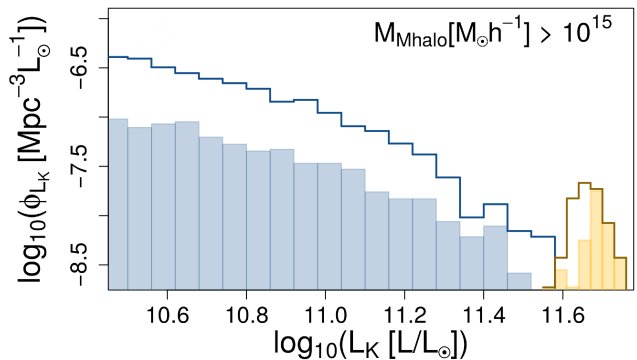}\\
\caption{Analogue to Figure\,\ref{fl11} but considering
halos more massive than $10^{12}\,{\rm M_{\odot}}$ (case C).}    
\label{fl12}    
\end{figure}

The restrictive choice for $V_{\rm R,max}$ for case (C) limits the 
identification of a significant portion of real
satellites for massive main halos (right panel of Figure\,\ref{frac212}),
comparable with massive clusters of galaxies, which present radial
velocity dispersions in the order of several hundreds km\,s$^{-1}$.
This is particularly striking for masses above $10^{15}\,{\rm M_{\odot}\,h^{-1}}$
whose conditional luminosity functions are presented in Figure\,\ref{fl12}.
As it can be seen from the figure, it leads to
a paucity of faint main halos and a 
shallower distribution for satellites, which result in ratios of 
$\approx 0.6$ and $\approx 0.3$, respectively. Therefore, the method only
associates half the number of bounded satellites to halos more massive 
than $10^{15}\,{\rm M_{\odot}\,h^{-1}}$. This paucity is mitigated for halos in
the bin $10^{14} < M_{\rm vir} < 10^{15}\,{\rm M_{\odot}\,h^{-1}}$ 
and it is absent for the
less massive bin where the ratios for main and satellite halos are
$\approx 1$ and $\approx 1.5$, respectively, similar to those in cases
(B) and (C).

\subsection{Dependence on approximations}

To clarify previous analysis it is important to take into consideration
how estimations ruled by observational limitations might affect the method.
In this issue we point to the effect that velocity uncertainties and 
the estimation of projected distances 
might yield differences in the method results. Thus, we 
repeated the application of the method on our sample but
considering two simplifications. First, 
to isolate the effect of the uncertainties in $V_{\rm R}$ we 
exclude them from the calculations.
As simulated uncertainties depend on the apparent $K$ magnitude, we 
restricted this analysis to halos less massive than $10^{12}\,{\rm M_{\odot}
\,h^{-1}}$, i.e. (fainter than $M_K = -23.1$\,mag). We split the sample
in three equally populated ranges of $V_{\rm R}$ and compared the fraction
of accurately classified halos as a function of their virial mass. No
significant changes were appreciated in the distributions, and the 
number of halos whose classification changed with respect to the
original one was $\approx 1\%$ in the three cases, but with a lower
fraction of success in the bin that corresponds to the lower $V_{\rm R}$ 
compared 
with the other ones. 
The negligible effect of this uncertainties in the overall analysis 
is expected due to the assumed
values for the linking-length parameter $V_{\rm R,max}$. We must notice
that the distribution of uncertainties assumed in Section\,\ref{vel.uncert}
implies for a galaxy with $K = 11$\,mag $eV_{\rm R} = 136\,{\rm km\,s^{-1}}$ 
at $3 \sigma$ level. 

To understand the different behaviour of the bin that corresponds to the
lower $V_{\rm R}$,
besides avoiding the uncertainties in $V_{\rm R}$, as a second approach we
calculate the projected distances for each pair of halos from the
real line-of-sight distance, instead of approaching it from the $V_{\rm R}$ 
mean (see Section \ref{proj.est}). The sample was also split in three
equally populated bins of $V_{\rm R}$. For neither the two furthest thirds, 
the comparison with the previous simplification results in no striking 
differences, with $\approx 3\%$ halos changing their classification.
There was an improvement of performance for the nearest bin, achieving a 
similar fraction of accurately classified halos than the other bins, 
with $\approx 13\%$ of the halos changing their classification in
comparison with the previous one. This points to the 
estimation of the line-of-sight distance as the main source of the 
discrepancy, due to the noise added by the relative peculiar velocity of 
the pair. This effect was mitigated for the other bins because the 
contribution from outflow velocities dominates the $V_{\rm R}$.

\subsection{Constraints to the method}

Even though our choices for linking-length 
parameters ensure a high fraction of 
success for both main and satellites halos, it is worth to go further 
with the analysis of typical projected distances. 

From Figure\,\ref{simul} we concluded that the mean distance from the 
main halos centre to their furthest satellite ($D_{Mh,max}$)
exceeds the typical virial radius for main halos more massive than 
$10^{13}\,{\rm M_{\odot}\,h^{-1}}$, evolving up to $\approx 1.4 R_{\rm vir}$.
Additionally, 
in the previous Section we noticed the differences in the distribution 
of projected distances to the main halo between real satellites and
fake positives (Figure\,\ref{proj}). For constraining the projected
distances, we select as an upper limit the 95th percentile of 
$D_{Mh,max}$, which barely matches the $2\sigma$ deviation from 
the mean value of $D_{\rm Mh,max}$ across the entire mass range. 
The curve can be accurately approximated by the function
\begin{equation}
{\rm D_{Mh,max}}(M) = \alpha \left( 1 + \frac{\beta 
(M-\gamma)}{\sqrt{1 + \beta^2 (M-\gamma)^2}} \right),
\end{equation}
where $M$ corresponds to the $M_{\rm vir}$ of the main halo in 
logarithmic scale and units of ${\rm M_{\odot}\,h^{-1}}$, 
and ${\rm D_{Mh,max}}$ is in units of ${\rm Mpc\,h^{-1}}$.
The fit of the function gives $\alpha= 2.66\pm0.11$, 
$\beta= -0.75\pm0.01$ and $\gamma= 14.93\pm0.05$.

An equivalent analysis can be done for the radial velocity 
differences between the main halos and their satellites. We fitted
the same function to the 95th percentile of ${\rm \Delta V_{R,max}}$ 
across the entire mass range, in units of ${\rm km\,s^{-1}}$. In this case, 
the parameters were $\alpha= 1255\pm19$, $\beta= -0.83\pm0.01$ 
and $\gamma= 14.03\pm0.03$.

\begin{figure}    
\includegraphics[width=80mm]{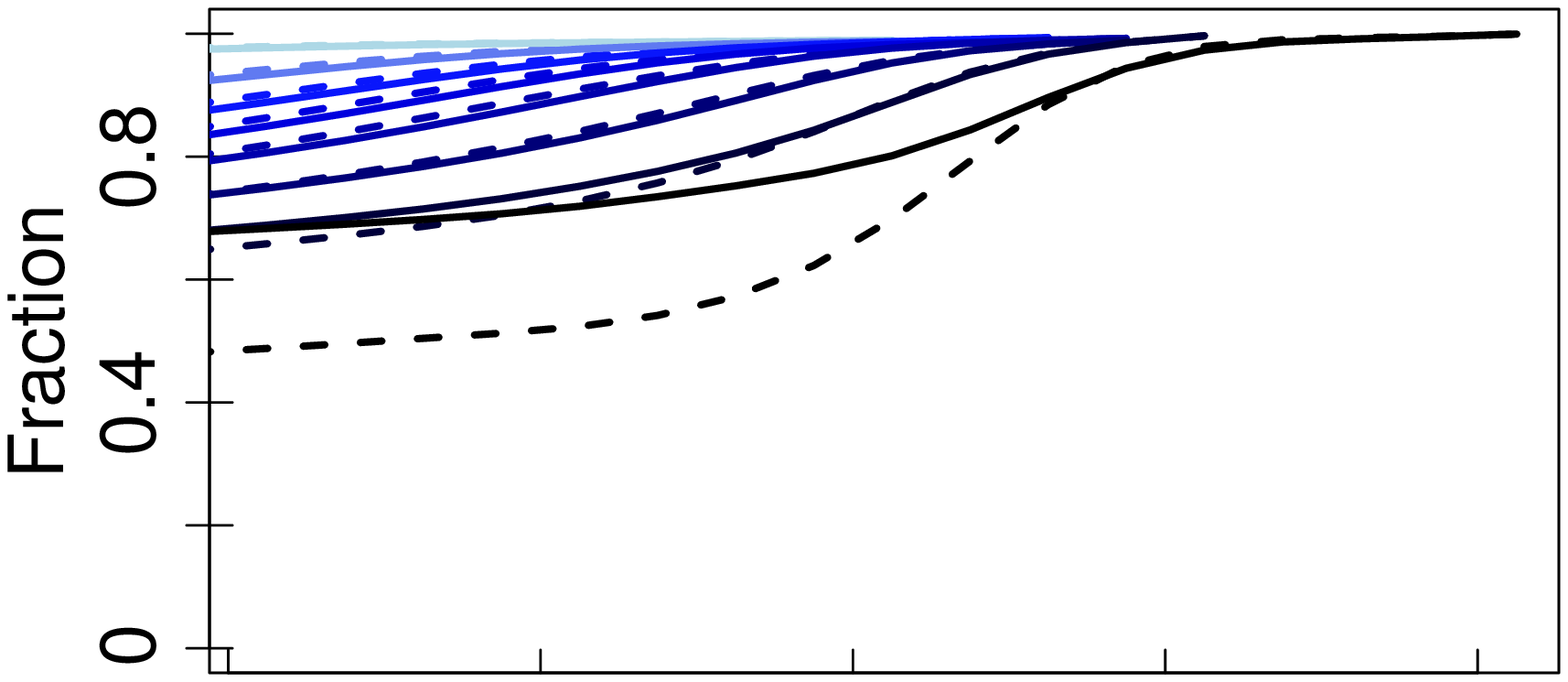}\\
\includegraphics[width=80mm]{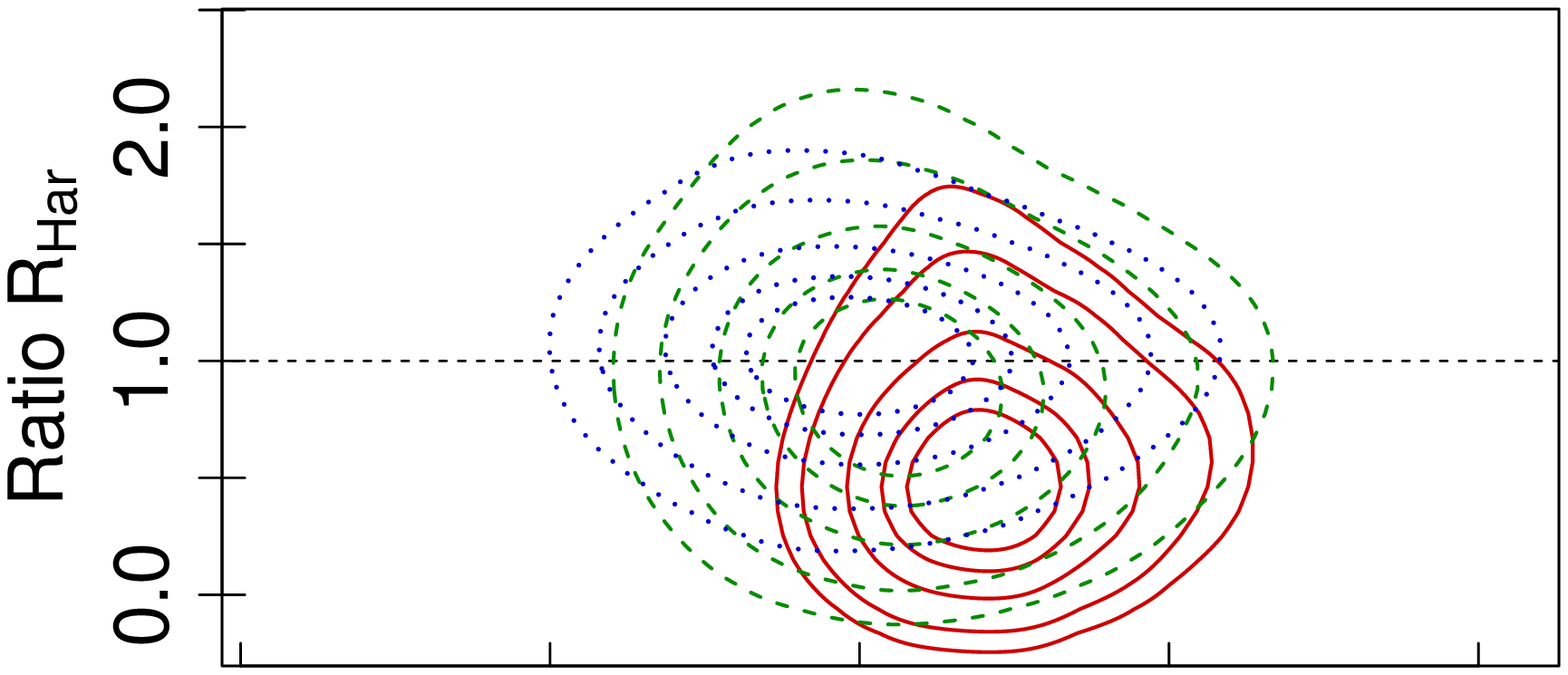}\\
\includegraphics[width=80mm]{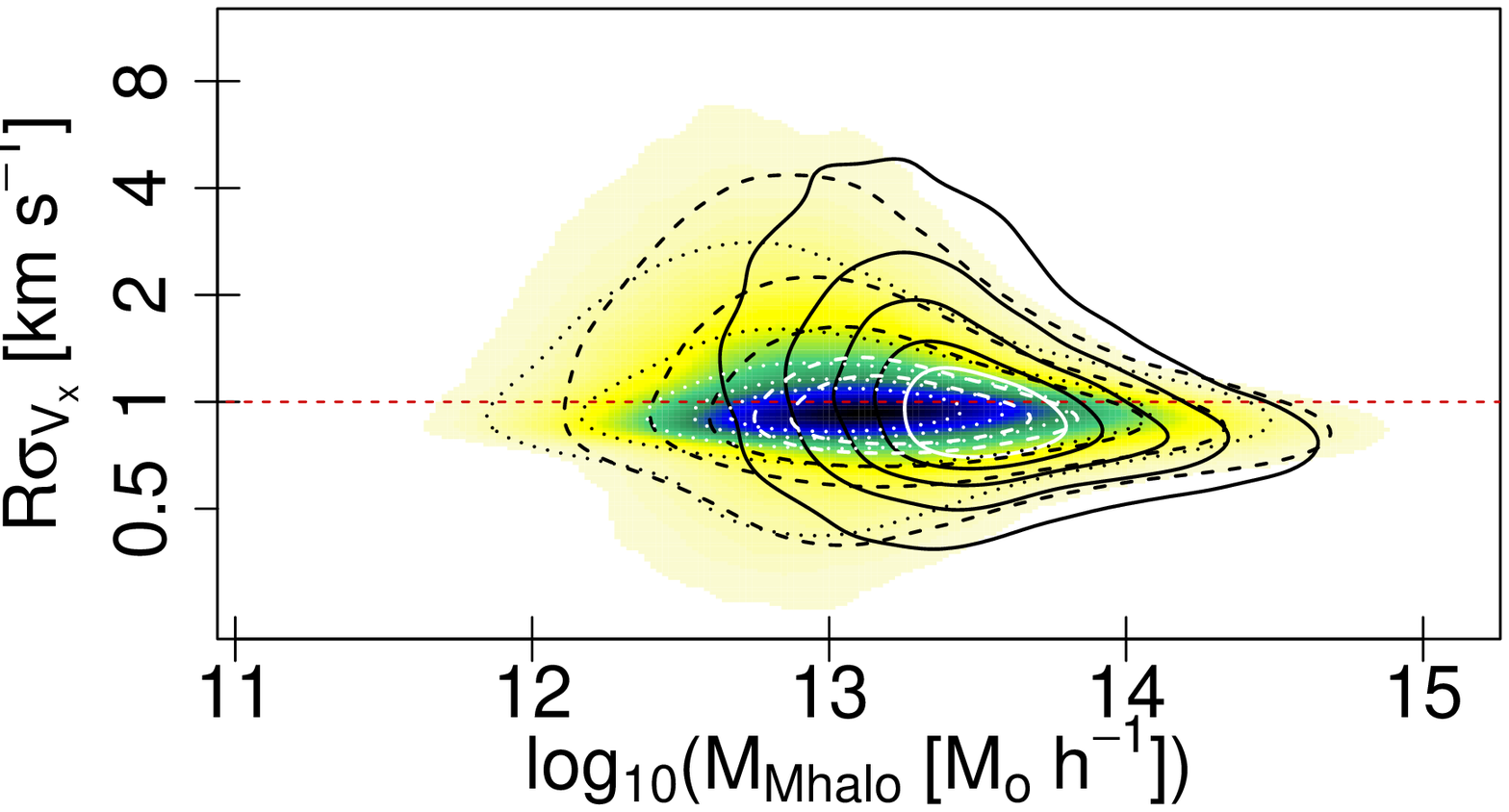}\\
\caption{{\bf Upper panel:} Fraction of halos (solid curves) and 
main halos (dashed curves) accurately classified as a function of
their main halo virial mass, as in Figure\,\ref{frac111} (case A). 
The sample was split in several ranges of environmental density, 
with colours becoming darker towards denser environments. {\bf Middle
panel:} Ratio between the harmonic radius (${\rm R_{Har}}$) of
halos associated by the method and halos gravitationally
bounded in case (A). The contour curves represent three bins of 
$R_{\rm S/O}$, as in Figure\,\ref{proj2}. {\bf Lower panel:}
Ratio between the velocity dispersion (R$\sigma_{\rm V_x}$) derived 
from halos associated by the method and real satellite halos, 
for the case (A). The colour gradient indicates the density
distribution, and the contour curves represent three bins of 
$R_{\rm S/O}$, as in Figure\,\ref{disp1}.}
\label{improve}    
\end{figure}

We rerun the percolation algorithm for halos more massive than 
$10^{11}\,{\rm M_{\odot}\,h^{-1}}$ (case A), applying the restrictions
in the maximum projected distance to the main halo ($D_{\rm Mh,max}$)
and the maximum difference in radial velocity (${\rm \Delta V_{R,max}}$).
These restrictions
resulted in an improvement in the fraction of halos accurately 
classified by the method. As an example, 
the upper panel of Figure\,\ref{improve} shows the evolution
of the fraction of halos (solid curves) and main halos (dashed 
curves) accurately classified as a function of their main halo 
virial mass, for several ranges of environmental density. This
is analogue to Figure\,\ref{frac111}, corresponding to case (A).
The middle panel shows the ratio of the harmonic radius
for halos associated by the method to halos gravitationally bounded
in case (A), representing an improvement in comparison with the
first row in Figure\,\ref{proj2}. The lower panel corresponds
to the ratio of velocity dispersions, analogue to Figure\,\ref{disp1}.
In this case the overall distribution follows the contour curves
corresponding to the bin $0.66< R_{\rm S/O} < 1$, pointing to
the minor relevance of the other two bins.

\begin{figure}    
\includegraphics[clip, width=80mm]{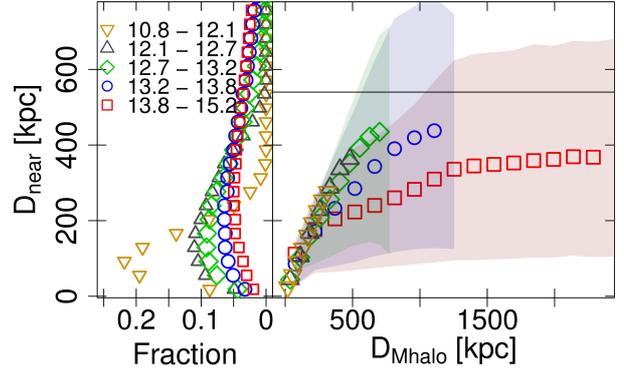}\\
\caption{{\bf Left panel:} Distribution of projected distances of 
satellite halos to their nearest neighbour, separated in five 
bins of mass equally populated. {\bf Right panel:} change in the
mean value of $D_{\rm near}$ for satellite halos, as a function of
the distance to the centre of their main halo $D_{\rm Mhalo}$. Shaded
regions represent the dispersion for each distribution.}    
\label{ddproj1}    
\end{figure}

In order to assess additional constraints, we also analysed the 
distribution of projected distances of satellite
halos to their nearest neighbour (i.e., the nearest halo that belongs to 
the same main halo). We split the sample of main halos in five mass bins,
each one equally populated of satellites halos. In each case, we rejected
the cases which do not fulfil the linking-length criteria for $V_{\rm R}$. 
This is shown in the left panel of Figure\,\ref{ddproj1} with different 
symbols. The labels refer to $M_{\rm vir}$ for the main halos in units of 
${\rm M_{\odot}\,h^{-1}}$ and logarithmic scale. 
The distributions present a similar behaviour, 
becoming more disperse for more massive halos. The projected distance 
corresponding to the 95-percentile evolve from 240\,kpc to 610\,kpc 
from the less massive to the most massive bins, with marginal changes
in the $D_{\rm near}$ distribution for main halos more massive than 
$10^{13}\,{\rm M_{\odot}\,h^{-1}}$. The latter values for the 95-percentile
are barely larger than $D_{\rm p,max}=525$\,kpc, the linking-length limit
chosen in Section\,\ref{sec.param} for halos more massive than 
$10^{11}\,\textrm{M}_\odot$. A similar result is obtained when halos
more massive than $10^{12}\,\textrm{M}_\odot$ are considered. Hence, this
might be used to select an appropriate linking-length parameter $D_{\rm p,max}$ 
in observational studies.

The right panel of Figure\,\ref{ddproj1} shows the change in the
mean value of $D_{\rm near}$ for satellite halos, as a function of
their projected distance to the centre of the corresponding main 
halo $D_{\rm Mhalo}$. 
The mass bins correspond to those indicated in the left panel,
keeping the same symbol references. For main halos with masses below 
$10^{13}\,{\rm M_{\odot}\,h^{-1}}$ the change of $D_{\rm near}$ as a function
of $D_{\rm Mhalo}$ does not depend on $M_{\rm vir}$.
Instead, from that value it starts to become
less steepy with increasing $M_{\rm vir}$. The filled regions
indicate the ranges $D_{\rm near}$ spans within
the 5th percentile lower and upper tails. 
The black solid line shows the linking-length parameter 
($D_{\rm p,max}$) chosen in Section\,\ref{permet}. 
The latter value is smaller than the 95th percentile for distances
to the main halo comparable with the $R_{\rm Vir}$, but in Fig\,\ref{proj}
it was found that the contribution of satellites at similar distances
is minor, hence its 95th percentile is barely negligible. 
A similar result is obtained when only halos more massive than 
$10^{12}\,{\rm M_{\odot}\,h^{-1}}$ are considered. In this case, the
typical number of satellites is lower, and the behaviour of 
 $D_{\rm near}$ do not vary with the virial mass of the main halo.
Then, no trend between the main halo virial mass and the typical value of
$D_{\rm near}$, e.g. its 95th percentile, is evident for halos 
more massive than $10^{13}\,{\rm M_{\odot}\,h^{-1}}$. Similar results
were obtained when the difference in radial velocities, 
${\rm \Delta V_{R}}$, was taken into account. Then, we avoided to 
apply variable linking-length parameters.

From the results in previous Sections we can conclude that the method 
is less accurate for dense environments and low mass halos. 
This is particularly noticeable for main halos, due to the large number
of field halos in the low mass regime. For instance, 
we selected from the simulation main halos in dense environments with 
more than ten satellites and virial masses from a few times $10^{13}$
to $10^{14}\,{\rm M_{\odot}}$, which correspond to the mass range of nearby
clusters of galaxies like Fornax and Virgo. For those to whom the
algorithm associated a larger number of halos than real satellites
(corresponding to $\approx 70$ per cent of the total),
the mean number of wrongly classified halos was $\approx 8.3$, with
$\approx 81$ per cent of them being field halos and only $\approx 5$
per cent belonging to systems with more than five members. From 
these fake positives, $\approx 85$ per cent present masses below
$10^{12}\,{\rm M_{\odot}}$. Consequently, the contribution to the virial 
mass of the main halo is negligible, but their inclusion might
lead to inaccurate mass estimations due to variations in the dispersion
of $V_{R\rm }$ (see Fig.\,\ref{disp1}).

We explored several options to identify main halos presenting a large 
fraction of fake positives, like Gaussian Mixture Model \citep{mur10} 
and nearest neighbours tests \citep{col96}, but the large fraction
of fake positives corresponding to field halos blur structures in
$V_{\rm R}$-space in the majority of the cases. A simplistic but 
efficient way to estimate the degree of contaminants in a cluster
catalogue is to compare the distribution of projected distances to
the main halo centre with the expected one, that presents a similar
behaviour at all virial masses ranges when it is defined in terms
of the main halo virial radius (see Figure\,\ref{proj}).
A particular case might be the analogue to a low-mass group
classified by the method as part of a more massive cluster of 
galaxies. To present analogues to nearby clusters of galaxies, we 
selected again main halos in dense environments, with more than ten 
satellites and virial masses from a few times $10^{13}$ to 
$10^{14}\,{\rm M_{\odot}}$. We focused on those presenting at least 
five fake positives corresponding to the same main halo. 
We run the substructure test described in \citet{col96}
for the ten nearest neighbours. 
In this test the statistic 
is ruled by the probability of the K-S two-sample distribution 
\citep{kol33,smi48}, between each galaxy and its nearest
neighbours, and the entire sample. The significance 
of the statistic is estimated by Monte Carlo simulations, in which 
the velocities of the galaxies are shuffled randomly. Hence, the 
zero hypothesis is that there is no substructure in the sample of 
galaxies. In a third of the main halos the 
presence of substructure cannot be ruled out at the 0.1 confidence 
level, and the proportion increases to nearly 50 per cent for the
0.2 confidence level. A control sample was chosen from the main 
halos in similar environments and mass range, but presenting a 
fraction of $R_{\rm S/O}$ from 0.9 to 1.1. From 
these, in 1\,per cent of the cases the presence of substructure 
cannot be ruled out at the 0.1 confidence level. Hence, although the 
percolation algorithm results in an overpopulation of satellites 
for massive halos, subsequent analysis might improve the characterization. 

In order to avoid adding noise from empirical fits and theoretical
assumptions, the resulted fits
were presented so far in terms of the halos
virial mass instead of parameters like the stellar mass or the 
luminosity which might allow a direct comparison from observations.
Despite of this, we are aware that observational surveys usually 
lack of accurate virial mass determinations due to the costs in telescope
time of this type of analysis. Accurate constraints of mass-to-light ratios
should be used to properly compare the results from the MDPL2 
simulation with observational surveys, but this exceeds the goals 
from the present paper.

\subsection{Implementation on observational data}

Although the build-up of a group/clusters catalogue from observational
surveys exceed the goals of this paper, it is worth testing the results 
with a sample of galaxies. We chose the surroundings of the Virgo cluster,
because it has been extensively studied in the literature, including its
substructure and satellite galaxies that might be falling in it.

The catalogue was obtained from the SDSS Data Release\,7 \citep{aba09},
which has been used in previous studies focused on the Virgo cluster
\citep{kim14,lis18}. The chosen region spans $30\times30$\,deg$^2$,
presenting galaxies that satisfy $172 < RA < 202$\,deg and $-5 < DE < 25$\,deg
and ${\rm V_R < 3000 km\,s^{-1}}$, that includes the Virgo
cluster and several groups of galaxies conforming the extended Virgo
cluster catalogue \citep[EVCC][]{kim14,kim16}. The SDSS spectroscopic 
survey covers galaxies brighter than $r \lesssim 17.77$, 
but bright and bulge-dominated galaxies were excluded due saturation 
problems. Hence, the catalogue was complemented with data from 
HyperLeda\footnote{http://leda.univ-lyon1.fr/} database \citep{mak14}. 
The distance modulus for each galaxy was calculated from their radial
velocities. This quantity was latter used to obtain absolute magnitudes
in the $K$ filter. From the MDPL2 simulation we found that a halo 
with $10^{11}\,{\rm M_{\odot}}$ presents $M_K \approx -21$, then we 
used this magnitude as a selection limit for the sample of galaxies, 
resulting in 305 galaxies.

\begin{figure}    
\includegraphics[clip, width=80mm,angle=270]{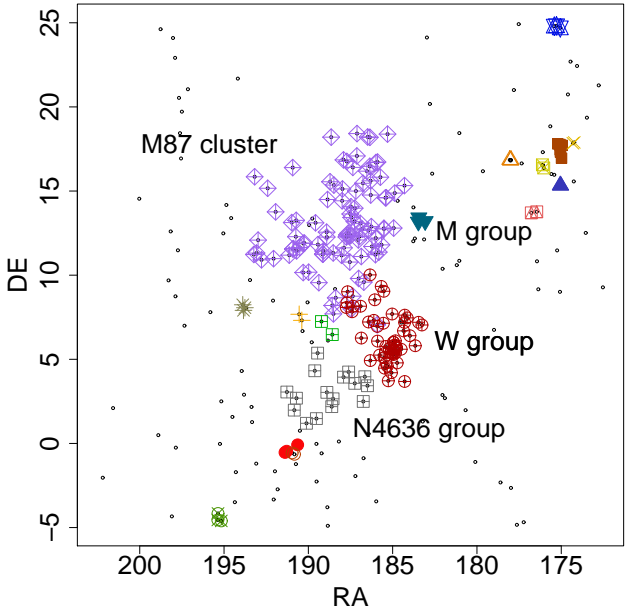}\\
\caption{Results from applying the percolation method with the linking 
length parameters calculated in Section\,{sec.param} for a sample of
galaxies brighter than $M_K=-21$ in the surroundings of the Virgo cluster.
Different symbols correspond to associations of galaxies found by the 
method.}    
\label{virgo}    
\end{figure}

We applied the percolation method to the sample, using the linking 
length parameters calculated in Section\,\ref{sec.param} for the 
indicated mass limit corresponding to the case~(A). The largest group 
of galaxies associated by the method might be identified with the main 
body of the Virgo cluster. This contains the giant ellipticals M\,87 
and M\,49, as well as other 90 members (blue violet diamonds in 
Figure\,\ref{virgo}). Although it should be stressed that the use 
of radial velocities to estimate the distance in the line-of-sight 
leads to NGC\,4501 as the most luminous galaxy of this association, 
instead of M\,87. For instance, this approach estimates for 
NGC\,4501 a distance modulus of $m_M\approx32.6$, while SNIa measurements 
results in $m-M\approx30.9$ \citep{man11} and Tully-Fisher relations 
in $m-M\approx31.5$ \citep{tul13}. Because redshift-independent 
measurements are not usually available in galaxy surveys, alternative
estimations should be introduced in order to minimise these uncertainties.

The groups of galaxies typically called the
W and M clouds were identified as independent associations to the Virgo
cluster, which is supported by discussion about their stage of infall
in literature \citep[e.g.][]{bin93,gav99,kim14,lis18}. The percolation
method also found the NGC\,4636 group, a dinamically old group, lacking 
of late-type galaxies \citep{kim14} and that exhibits intragroup 
X-ray emission that is distinct from the emission of the Virgo cluster
\citep{brou06}. 

\section{Summary} 

We applied a percolation method, commonly used in observational
astronomy to identify groups and clusters, to the catalogue of 
halos corresponding to the local Universe ($z=0$) from the 
MDPL2 cosmological dark matter simulation. Two different cuts
in halo mass were assumed in order to reproduce results for
different completeness depths. We added noise to our incoming 
parameters, projected distances and radial velocities, looking 
to reproduce the limitations from an observational surveys.
We summarizes our findings as follows: 
\begin{itemize}
\item The method is less efficient in dense environments, 
presenting a large fraction of fake positives. The uncertainties 
in the incoming parameters are not responsible for the 
misclassified halos.
\item The selection of the linking-length parameters is crucial
for the success of the method. Because the number of satellite
halos increases strongly with virial mass, the parameters that
produce better results are defined by the most massive main 
halos. 
\item The linking-length parameter in projected distance 
depends strongly on the mass cut of the analysis, and it
might be approximated by the 95-percentile of the distribution
of projected distances to the closest neighbour in massive
main halos.
\item Observational properties derived from contaminated
catalogues, like velocity dispersions or luminosity functions
might deviate significantly from the accurate ones.
\item Constraints to maximum projected distance and difference
in radial velocities, in terms of the main halo virial mass,
can contribute to significantly reduce fraction of misclassified
halos.
\item The analysis of the distribution of projected distances
might be helpful to identify largely contaminated systems. The
use of substructure tests can also provide information in 
particular cases.
\item The method with the constraints derived in Section\,4.6
was applied to a sample of galaxies in the surroundings of the Virgo
cluster with halo masses above $10^{11}\,{\rm M_{\odot}}$. The
groups identified by the method are in agreement with the 
results from the literature for this region.
\end{itemize}

\section*{Acknowledgements}
This research was 
funded with grants from Consejo Nacional de Investigaciones   
Cient\'{\i}ficas y T\'ecnicas de la Rep\'ublica Argentina (PIP 112-201101-00393), 
Agencia Nacional de Promoci\'on Cient\'{\i}fica y Tecnol\'ogica (PICT-2013-0317), 
and Universidad Nacional de La Plata (UNLP 11-G124), Argentina.
CVM acknowledges financial support from the Max Planck Society 
through a Partner Group grant.

\bibliographystyle{mnras}
\bibliography{biblio}

\begin{thebibliography}{}
\makeatletter
\relax
\def\mn@urlcharsother{\let\do\@makeother \do\$\do\&\do\#\do\^\do\_\do\%\do\~}
\def\mn@doi{\begingroup\mn@urlcharsother \@ifnextchar [ {\mn@doi@}
  {\mn@doi@[]}}
\def\mn@doi@[#1]#2{\def\@tempa{#1}\ifx\@tempa\@empty \href
  {http://dx.doi.org/#2} {doi:#2}\else \href {http://dx.doi.org/#2} {#1}\fi
  \endgroup}
\def\mn@eprint#1#2{\mn@eprint@#1:#2::\@nil}
\def\mn@eprint@arXiv#1{\href {http://arxiv.org/abs/#1} {{\tt arXiv:#1}}}
\def\mn@eprint@dblp#1{\href {http://dblp.uni-trier.de/rec/bibtex/#1.xml}
  {dblp:#1}}
\def\mn@eprint@#1:#2:#3:#4\@nil{\def\@tempa {#1}\def\@tempb {#2}\def\@tempc
  {#3}\ifx \@tempc \@empty \let \@tempc \@tempb \let \@tempb \@tempa \fi \ifx
  \@tempb \@empty \def\@tempb {arXiv}\fi \@ifundefined
  {mn@eprint@\@tempb}{\@tempb:\@tempc}{\expandafter \expandafter \csname
  mn@eprint@\@tempb\endcsname \expandafter{\@tempc}}}

\bibitem[\protect\citeauthoryear{{Abazajian} et~al.,}{{Abazajian}
  et~al.}{2009}]{aba09}
{Abazajian} K.~N.,  et~al., 2009, \mn@doi [\apjs]
  {10.1088/0067-0049/182/2/543}, \href
  {http://adsabs.harvard.edu/abs/2009ApJS..182..543A} {182, 543}

\bibitem[\protect\citeauthoryear{{Baldry}, {Balogh}, {Bower}, {Glazebrook},
  {Nichol}, {Bamford}  \& {Budavari}}{{Baldry} et~al.}{2006}]{bal06}
{Baldry} I.~K.,  {Balogh} M.~L.,  {Bower} R.~G.,  {Glazebrook} K.,  {Nichol}
  R.~C.,  {Bamford} S.~P.,   {Budavari} T.,  2006, \mn@doi [\mnras]
  {10.1111/j.1365-2966.2006.11081.x}, \href
  {http://adsabs.harvard.edu/abs/2006MNRAS.373..469B} {373, 469}

\bibitem[\protect\citeauthoryear{{Behroozi}, {Wechsler}  \& {Wu}}{{Behroozi}
  et~al.}{2013}]{beh13}
{Behroozi} P.~S.,  {Wechsler} R.~H.,   {Wu} H.-Y.,  2013, \mn@doi [ApJ]
  {10.1088/0004-637X/762/2/109}, \href
  {https://ui.adsabs.harvard.edu/abs/2013ApJ...762..109B} {762, 109}

\bibitem[\protect\citeauthoryear{{Berlind} et~al.,}{{Berlind}
  et~al.}{2006}]{ber06}
{Berlind} A.~A.,  et~al., 2006, \mn@doi [\apjs] {10.1086/508170}, \href
  {http://adsabs.harvard.edu/abs/2006ApJS..167....1B} {167, 1}

\bibitem[\protect\citeauthoryear{{Binggeli}, {Popescu}  \&
  {Tammann}}{{Binggeli} et~al.}{1993}]{bin93}
{Binggeli} B.,  {Popescu} C.~C.,   {Tammann} G.~A.,  1993, \aaps, \href
  {https://ui.adsabs.harvard.edu/abs/1993A&AS...98..275B} {98, 275}

\bibitem[\protect\citeauthoryear{{Boylan-Kolchin}, {Springel}, {White},
  {Jenkins}  \& {Lemson}}{{Boylan-Kolchin} et~al.}{2009}]{boy09}
{Boylan-Kolchin} M.,  {Springel} V.,  {White} S.~D.~M.,  {Jenkins} A.,
  {Lemson} G.,  2009, \mn@doi [\mnras] {10.1111/j.1365-2966.2009.15191.x},
  \href {http://adsabs.harvard.edu/abs/2009MNRAS.398.1150B} {398, 1150}

\bibitem[\protect\citeauthoryear{{Brough}, {Forbes}, {Kilborn}  \&
  {Couch}}{{Brough} et~al.}{2006}]{brou06}
{Brough} S.,  {Forbes} D.~A.,  {Kilborn} V.~A.,   {Couch} W.,  2006, \mn@doi
  [\mnras] {10.1111/j.1365-2966.2006.10542.x}, \href
  {https://ui.adsabs.harvard.edu/abs/2006MNRAS.370.1223B} {370, 1223}

\bibitem[\protect\citeauthoryear{{Caso}, {Bassino}  \& {G{\'o}mez}}{{Caso}
  et~al.}{2015}]{cas15a}
{Caso} J.~P.,  {Bassino} L.~P.,   {G{\'o}mez} M.,  2015, \mn@doi [MNRAS]
  {10.1093/mnras/stv2015}, \href
  {http://adsabs.harvard.edu/abs/2015MNRAS.453.4421C} {453, 4421}

\bibitem[\protect\citeauthoryear{{Colless} \& {Dunn}}{{Colless} \&
  {Dunn}}{1996}]{col96}
{Colless} M.,  {Dunn} A.~M.,  1996, \mn@doi [ApJ] {10.1086/176827}, \href
  {http://adsabs.harvard.edu/abs/1996ApJ...458..435C} {458, 435}

\bibitem[\protect\citeauthoryear{{Conroy}, {Wechsler}  \& {Kravtsov}}{{Conroy}
  et~al.}{2006}]{con06}
{Conroy} C.,  {Wechsler} R.~H.,   {Kravtsov} A.~V.,  2006, \mn@doi [\apj]
  {10.1086/503602}, \href {http://adsabs.harvard.edu/abs/2006ApJ...647..201C}
  {647, 201}

\bibitem[\protect\citeauthoryear{{Cora} et~al.,}{{Cora} et~al.}{2018}]{cor18}
{Cora} S.~A.,  et~al., 2018, \mn@doi [\mnras] {10.1093/mnras/sty1131}, \href
  {http://adsabs.harvard.edu/abs/2018MNRAS.479....2C} {479, 2}

\bibitem[\protect\citeauthoryear{{Crook}, {Huchra}, {Martimbeau}, {Masters},
  {Jarrett}  \& {Macri}}{{Crook} et~al.}{2007}]{cro07}
{Crook} A.~C.,  {Huchra} J.~P.,  {Martimbeau} N.,  {Masters} K.~L.,  {Jarrett}
  T.,   {Macri} L.~M.,  2007, \mn@doi [\apj] {10.1086/510201}, \href
  {http://adsabs.harvard.edu/abs/2007ApJ...655..790C} {655, 790}

\bibitem[\protect\citeauthoryear{{Duarte} \& {Mamon}}{{Duarte} \&
  {Mamon}}{2014}]{dua14}
{Duarte} M.,  {Mamon} G.~A.,  2014, \mn@doi [\mnras] {10.1093/mnras/stu378},
  \href {http://adsabs.harvard.edu/abs/2014MNRAS.440.1763D} {440, 1763}

\bibitem[\protect\citeauthoryear{{Eke} et~al.,}{{Eke} et~al.}{2004a}]{eke04a}
{Eke} V.~R.,  et~al., 2004a, \mn@doi [\mnras]
  {10.1111/j.1365-2966.2004.07408.x}, \href
  {http://adsabs.harvard.edu/abs/2004MNRAS.348..866E} {348, 866}

\bibitem[\protect\citeauthoryear{{Eke} et~al.,}{{Eke} et~al.}{2004b}]{eke04b}
{Eke} V.~R.,  et~al., 2004b, \mn@doi [\mnras]
  {10.1111/j.1365-2966.2004.08354.x}, \href
  {http://adsabs.harvard.edu/abs/2004MNRAS.355..769E} {355, 769}

\bibitem[\protect\citeauthoryear{{Finn} et~al.,}{{Finn} et~al.}{2018}]{fin18}
{Finn} R.~A.,  et~al., 2018, \mn@doi [\apj] {10.3847/1538-4357/aac32a}, \href
  {http://adsabs.harvard.edu/abs/2018ApJ...862..149F} {862, 149}

\bibitem[\protect\citeauthoryear{{Firth}, {Evstigneeva}, {Jones}, {Drinkwater},
  {Phillipps}  \& {Gregg}}{{Firth} et~al.}{2006}]{fir06}
{Firth} P.,  {Evstigneeva} E.~A.,  {Jones} J.~B.,  {Drinkwater} M.~J.,
  {Phillipps} S.,   {Gregg} M.~D.,  2006, \mn@doi [\mnras]
  {10.1111/j.1365-2966.2006.10993.x}, \href
  {http://adsabs.harvard.edu/abs/2006MNRAS.372.1856F} {372, 1856}

\bibitem[\protect\citeauthoryear{{Garcia}}{{Garcia}}{1993}]{gar93}
{Garcia} A.~M.,  1993, A\&AS, \href
  {http://adsabs.harvard.edu/abs/1993A%26AS..100...47G} {100, 47}

\bibitem[\protect\citeauthoryear{{Gavazzi} \& {Boselli}}{{Gavazzi} \&
  {Boselli}}{1996}]{gav96}
{Gavazzi} G.,  {Boselli} A.,  1996, Astrophysical Letters and Communications,
  \href {http://adsabs.harvard.edu/abs/1996ApL%26C..35....1G} {35, 1}

\bibitem[\protect\citeauthoryear{{Gavazzi}, {Boselli}, {Scodeggio}, {Pierini}
  \& {Belsole}}{{Gavazzi} et~al.}{1999}]{gav99}
{Gavazzi} G.,  {Boselli} A.,  {Scodeggio} M.,  {Pierini} D.,   {Belsole} E.,
  1999, \mn@doi [\mnras] {10.1046/j.1365-8711.1999.02350.x}, \href
  {https://ui.adsabs.harvard.edu/abs/1999MNRAS.304..595G} {304, 595}

\bibitem[\protect\citeauthoryear{{Geller}, {Diaferio}  \& {Kurtz}}{{Geller}
  et~al.}{1999}]{gel99}
{Geller} M.~J.,  {Diaferio} A.,   {Kurtz} M.~J.,  1999, \mn@doi [\apjl]
  {10.1086/312024}, \href {http://adsabs.harvard.edu/abs/1999ApJ...517L..23G}
  {517, L23}

\bibitem[\protect\citeauthoryear{{Gonzalez-Perez}, {Lacey}, {Baugh}, {Lagos},
  {Helly}, {Campbell}  \& {Mitchell}}{{Gonzalez-Perez} et~al.}{2014}]{gon14}
{Gonzalez-Perez} V.,  {Lacey} C.~G.,  {Baugh} C.~M.,  {Lagos} C.~D.~P.,
  {Helly} J.,  {Campbell} D.~J.~R.,   {Mitchell} P.~D.,  2014, \mn@doi [\mnras]
  {10.1093/mnras/stt2410}, \href
  {http://adsabs.harvard.edu/abs/2014MNRAS.439..264G} {439, 264}

\bibitem[\protect\citeauthoryear{{Hern{\'a}ndez-Toledo}, {V{\'a}zquez-Mata},
  {Mart{\'{\i}}nez-V{\'a}zquez}, {Avila Reese}, {M{\'e}ndez-Hern{\'a}ndez},
  {Ortega-Esbr{\'{\i}}}  \& {N{\'u}{\~n}ez}}{{Hern{\'a}ndez-Toledo}
  et~al.}{2008}]{her08}
{Hern{\'a}ndez-Toledo} H.~M.,  {V{\'a}zquez-Mata} J.~A.,
  {Mart{\'{\i}}nez-V{\'a}zquez} L.~A.,  {Avila Reese} V.,
  {M{\'e}ndez-Hern{\'a}ndez} H.,  {Ortega-Esbr{\'{\i}}} S.,   {N{\'u}{\~n}ez}
  J.~P.~M.,  2008, \mn@doi [\aj] {10.1088/0004-6256/136/5/2115}, \href
  {http://adsabs.harvard.edu/abs/2008AJ....136.2115H} {136, 2115}

\bibitem[\protect\citeauthoryear{{Hess}, {Jarrett}, {Carignan}, {Passmoor}  \&
  {Goedhart}}{{Hess} et~al.}{2015}]{hes15}
{Hess} K.~M.,  {Jarrett} T.~H.,  {Carignan} C.,  {Passmoor} S.~S.,   {Goedhart}
  S.,  2015, preprint, \href
  {http://adsabs.harvard.edu/abs/2015arXiv150606143H} {} (\mn@eprint {arXiv}
  {1506.06143})

\bibitem[\protect\citeauthoryear{{Hirschmann}, {De Lucia}, {Iovino}  \&
  {Cucciati}}{{Hirschmann} et~al.}{2013}]{hir13}
{Hirschmann} M.,  {De Lucia} G.,  {Iovino} A.,   {Cucciati} O.,  2013, \mn@doi
  [\mnras] {10.1093/mnras/stt827}, \href
  {http://adsabs.harvard.edu/abs/2013MNRAS.433.1479H} {433, 1479}

\bibitem[\protect\citeauthoryear{{Huchra} \& {Geller}}{{Huchra} \&
  {Geller}}{1982}]{huc82}
{Huchra} J.~P.,  {Geller} M.~J.,  1982, \mn@doi [\apj] {10.1086/160000}, \href
  {http://adsabs.harvard.edu/abs/1982ApJ...257..423H} {257, 423}

\bibitem[\protect\citeauthoryear{{Huchra} et~al.,}{{Huchra}
  et~al.}{2012}]{huc12}
{Huchra} J.~P.,  et~al., 2012, \mn@doi [ApJS] {10.1088/0067-0049/199/2/26},
  \href {http://adsabs.harvard.edu/abs/2012ApJS..199...26H} {199, 26}

\bibitem[\protect\citeauthoryear{{Ivezi{\'c}} et~al.,}{{Ivezi{\'c}}
  et~al.}{2008}]{ive08}
{Ivezi{\'c}} {\v{Z}}.,  et~al., 2008, arXiv e-prints, \href
  {https://ui.adsabs.harvard.edu/\#abs/2008arXiv0805.2366I} {p.
  arXiv:0805.2366}

\bibitem[\protect\citeauthoryear{{Kawinwanichakij} et~al.,}{{Kawinwanichakij}
  et~al.}{2017}]{kaw17}
{Kawinwanichakij} L.,  et~al., 2017, \mn@doi [\apj] {10.3847/1538-4357/aa8b75},
  \href {http://adsabs.harvard.edu/abs/2017ApJ...847..134K} {847, 134}

\bibitem[\protect\citeauthoryear{{Kim} et~al.,}{{Kim} et~al.}{2014}]{kim14}
{Kim} S.,  et~al., 2014, \mn@doi [ApJS] {10.1088/0067-0049/215/2/22}, \href
  {http://adsabs.harvard.edu/abs/2014ApJS..215...22K} {215, 22}

\bibitem[\protect\citeauthoryear{{Kim} et~al.,}{{Kim} et~al.}{2016}]{kim16}
{Kim} S.,  et~al., 2016, \mn@doi [\apj] {10.3847/1538-4357/833/2/207}, \href
  {http://adsabs.harvard.edu/abs/2016ApJ...833..207K} {833, 207}

\bibitem[\protect\citeauthoryear{{Klypin}, {Trujillo-Gomez}  \&
  {Primack}}{{Klypin} et~al.}{2011}]{kly11}
{Klypin} A.~A.,  {Trujillo-Gomez} S.,   {Primack} J.,  2011, \mn@doi [\apj]
  {10.1088/0004-637X/740/2/102}, \href
  {http://adsabs.harvard.edu/abs/2011ApJ...740..102K} {740, 102}

\bibitem[\protect\citeauthoryear{Klypin, Yepes, Gottl\"ober, Prada  \&
  Heb}{Klypin et~al.}{2016}]{kly16}
Klypin A.,  Yepes G.,  Gottl\"ober S.,  Prada F.,   Heb S.,  2016, \mn@doi
  [MNRAS] {10.1093/mnras/stw248}, 457, 4340

\bibitem[\protect\citeauthoryear{{Knebe} et~al.,}{{Knebe} et~al.}{2011}]{kne11}
{Knebe} A.,  et~al., 2011, \mn@doi [MNRAS] {10.1111/j.1365-2966.2011.18858.x},
  \href {http://adsabs.harvard.edu/abs/2011MNRAS.415.2293K} {415, 2293}

\bibitem[\protect\citeauthoryear{{Knebe} et~al.,}{{Knebe} et~al.}{2018}]{kne18}
{Knebe} A.,  et~al., 2018, \mn@doi [\mnras] {10.1093/mnras/stx2662}, \href
  {http://adsabs.harvard.edu/abs/2018MNRAS.474.5206K} {474, 5206}

\bibitem[\protect\citeauthoryear{Kochanek et~al.,}{Kochanek
  et~al.}{2001}]{koc01}
Kochanek C.~S.,  et~al., 2001, \mn@doi [ApJ] {10.1086/322488}, 560, 566

\bibitem[\protect\citeauthoryear{{Kolmogorov}}{{Kolmogorov}}{1933}]{kol33}
{Kolmogorov} A.,  1933, {Giornale dell'Istituto Italiano degli Attuari}, 4, 83

\bibitem[\protect\citeauthoryear{{Kubo}, {Stebbins}, {Annis}, {Dell'Antonio},
  {Lin}, {Khiabanian}  \& {Frieman}}{{Kubo} et~al.}{2007}]{kub07}
{Kubo} J.~M.,  {Stebbins} A.,  {Annis} J.,  {Dell'Antonio} I.~P.,  {Lin} H.,
  {Khiabanian} H.,   {Frieman} J.~A.,  2007, \mn@doi [\apj] {10.1086/523101},
  \href {http://adsabs.harvard.edu/abs/2007ApJ...671.1466K} {671, 1466}

\bibitem[\protect\citeauthoryear{{Lacerna}, {Hern{\'a}ndez-Toledo},
  {Avila-Reese}, {Abonza-Sane}  \& {del Olmo}}{{Lacerna} et~al.}{2016}]{lac16}
{Lacerna} I.,  {Hern{\'a}ndez-Toledo} H.~M.,  {Avila-Reese} V.,  {Abonza-Sane}
  J.,   {del Olmo} A.,  2016, \mn@doi [\aap] {10.1051/0004-6361/201527844},
  \href {http://adsabs.harvard.edu/abs/2016A%26A...588A..79L} {588, A79}

\bibitem[\protect\citeauthoryear{{Lisker}, {Vijayaraghavan}, {Janz},
  {Gallagher}, {Engler}  \& {Urich}}{{Lisker} et~al.}{2018}]{lis18}
{Lisker} T.,  {Vijayaraghavan} R.,  {Janz} J.,  {Gallagher} III J.~S.,
  {Engler} C.,   {Urich} L.,  2018, \mn@doi [\apj] {10.3847/1538-4357/aadae1},
  \href {http://adsabs.harvard.edu/abs/2018ApJ...865...40L} {865, 40}

\bibitem[\protect\citeauthoryear{{{\L}okas} \& {Mamon}}{{{\L}okas} \&
  {Mamon}}{2003}]{lok03}
{{\L}okas} E.~L.,  {Mamon} G.~A.,  2003, \mn@doi [\mnras]
  {10.1046/j.1365-8711.2003.06684.x}, \href
  {http://adsabs.harvard.edu/abs/2003MNRAS.343..401L} {343, 401}

\bibitem[\protect\citeauthoryear{{Makarov} \& {Karachentsev}}{{Makarov} \&
  {Karachentsev}}{2011}]{mak11}
{Makarov} D.,  {Karachentsev} I.,  2011, \mn@doi [MNRAS]
  {10.1111/j.1365-2966.2010.18071.x}, \href
  {http://adsabs.harvard.edu/abs/2011MNRAS.412.2498M} {412, 2498}

\bibitem[\protect\citeauthoryear{{Makarov}, {Prugniel}, {Terekhova}, {Courtois}
   \& {Vauglin}}{{Makarov} et~al.}{2014}]{mak14}
{Makarov} D.,  {Prugniel} P.,  {Terekhova} N.,  {Courtois} H.,   {Vauglin} I.,
  2014, \mn@doi [\aap] {10.1051/0004-6361/201423496}, \href
  {http://adsabs.harvard.edu/abs/2014A%26A...570A..13M} {570, A13}

\bibitem[\protect\citeauthoryear{{Mandel}, {Narayan}  \& {Kirshner}}{{Mandel}
  et~al.}{2011}]{man11}
{Mandel} K.~S.,  {Narayan} G.,   {Kirshner} R.~P.,  2011, \mn@doi [\apj]
  {10.1088/0004-637X/731/2/120}, \href
  {https://ui.adsabs.harvard.edu/abs/2011ApJ...731..120M} {731, 120}

\bibitem[\protect\citeauthoryear{{Materne}}{{Materne}}{1978}]{mat78}
{Materne} J.,  1978, \aap, \href
  {http://adsabs.harvard.edu/abs/1978A%26A....63..401M} {63, 401}

\bibitem[\protect\citeauthoryear{{M{\'e}ndez}, {Teodorescu}, {Kudritzki}  \&
  {Burkert}}{{M{\'e}ndez} et~al.}{2009}]{men09}
{M{\'e}ndez} R.~H.,  {Teodorescu} A.~M.,  {Kudritzki} R.-P.,   {Burkert} A.,
  2009, \mn@doi [\apj] {10.1088/0004-637X/691/1/228}, \href
  {http://adsabs.harvard.edu/abs/2009ApJ...691..228M} {691, 228}

\bibitem[\protect\citeauthoryear{{Muratov} \& {Gnedin}}{{Muratov} \&
  {Gnedin}}{2010}]{mur10}
{Muratov} A.~L.,  {Gnedin} O.~Y.,  2010, \mn@doi [ApJ]
  {10.1088/0004-637X/718/2/1266}, \href
  {http://adsabs.harvard.edu/abs/2010ApJ...718.1266M} {718, 1266}

\bibitem[\protect\citeauthoryear{{Niemi}, {Hein{\"a}m{\"a}ki}, {Nurmi}  \&
  {Saar}}{{Niemi} et~al.}{2010}]{nie10}
{Niemi} S.-M.,  {Hein{\"a}m{\"a}ki} P.,  {Nurmi} P.,   {Saar} E.,  2010,
  \mn@doi [MNRAS] {10.1111/j.1365-2966.2010.16457.x}, \href
  {http://adsabs.harvard.edu/abs/2010MNRAS.405..477N} {405, 477}

\bibitem[\protect\citeauthoryear{{Paturel}}{{Paturel}}{1979}]{pat79}
{Paturel} G.,  1979, \aap, \href
  {http://adsabs.harvard.edu/abs/1979A%26A....71..106P} {71, 106}

\bibitem[\protect\citeauthoryear{{Planck Collaboration} et~al.,}{{Planck
  Collaboration} et~al.}{2013}]{pla13}
{Planck Collaboration} et~al., 2013, preprint, \href
  {http://adsabs.harvard.edu/abs/2013arXiv1303.5076P} {} (\mn@eprint {arXiv}
  {1303.5076})

\bibitem[\protect\citeauthoryear{{Rasmussen}, {Mulchaey}, {Bai}, {Ponman},
  {Raychaudhury}  \& {Dariush}}{{Rasmussen} et~al.}{2012}]{ras12}
{Rasmussen} J.,  {Mulchaey} J.~S.,  {Bai} L.,  {Ponman} T.~J.,  {Raychaudhury}
  S.,   {Dariush} A.,  2012, \mn@doi [\apj] {10.1088/0004-637X/757/2/122},
  \href {http://adsabs.harvard.edu/abs/2012ApJ...757..122R} {757, 122}

\bibitem[\protect\citeauthoryear{{Ricci} et~al.,}{{Ricci} et~al.}{2018}]{ric18}
{Ricci} M.,  et~al., 2018, \mn@doi [\aap] {10.1051/0004-6361/201832989}, \href
  {http://adsabs.harvard.edu/abs/2018A%26A...620A..13R} {620, A13}

\bibitem[\protect\citeauthoryear{{Richtler}, {Salinas}, {Lane}, {Hilker}  \&
  {Schirmer}}{{Richtler} et~al.}{2015}]{ric15}
{Richtler} T.,  {Salinas} R.,  {Lane} R.~R.,  {Hilker} M.,   {Schirmer} M.,
  2015, \mn@doi [A\&A] {10.1051/0004-6361/201424530}, \href
  {http://adsabs.harvard.edu/abs/2015A%26A...574A..21R} {574, A21}

\bibitem[\protect\citeauthoryear{{Robotham} et~al.,}{{Robotham}
  et~al.}{2011}]{rob11}
{Robotham} A.~S.~G.,  et~al., 2011, \mn@doi [\mnras]
  {10.1111/j.1365-2966.2011.19217.x}, \href
  {http://adsabs.harvard.edu/abs/2011MNRAS.416.2640R} {416, 2640}

\bibitem[\protect\citeauthoryear{{Salinas}, {Richtler}, {Bassino}, {Romanowsky}
   \& {Schuberth}}{{Salinas} et~al.}{2012}]{sal12}
{Salinas} R.,  {Richtler} T.,  {Bassino} L.~P.,  {Romanowsky} A.~J.,
  {Schuberth} Y.,  2012, \mn@doi [A\&A] {10.1051/0004-6361/201116517}, \href
  {http://adsabs.harvard.edu/abs/2012A%26A...538A..87S} {538, A87}

\bibitem[\protect\citeauthoryear{{Schaye} et~al.,}{{Schaye}
  et~al.}{2015}]{sch15}
{Schaye} J.,  et~al., 2015, \mn@doi [\mnras] {10.1093/mnras/stu2058}, \href
  {http://adsabs.harvard.edu/abs/2015MNRAS.446..521S} {446, 521}

\bibitem[\protect\citeauthoryear{{Schechter}}{{Schechter}}{1976}]{sch76}
{Schechter} P.,  1976, \mn@doi [\apj] {10.1086/154079}, \href
  {http://adsabs.harvard.edu/abs/1976ApJ...203..297S} {203, 297}

\bibitem[\protect\citeauthoryear{{Skrutskie} et~al.,}{{Skrutskie}
  et~al.}{2006}]{skr06}
{Skrutskie} M.~F.,  et~al., 2006, \mn@doi [\aj] {10.1086/498708}, \href
  {http://adsabs.harvard.edu/abs/2006AJ....131.1163S} {131, 1163}

\bibitem[\protect\citeauthoryear{{Smirnov}}{{Smirnov}}{1948}]{smi48}
{Smirnov} N.,  1948, {The Annals of Mathematical Statistics}, 19, 279

\bibitem[\protect\citeauthoryear{{Tal}, {van Dokkum}, {Nelan}  \&
  {Bezanson}}{{Tal} et~al.}{2009}]{tal09}
{Tal} T.,  {van Dokkum} P.~G.,  {Nelan} J.,   {Bezanson} R.,  2009, \mn@doi
  [AJ] {10.1088/0004-6256/138/5/1417}, \href
  {http://adsabs.harvard.edu/abs/2009AJ....138.1417T} {138, 1417}

\bibitem[\protect\citeauthoryear{{Tempel}, {Kruuse}, {Kipper}, {Tuvikene},
  {Sorce}  \& {Stoica}}{{Tempel} et~al.}{2018}]{tem18}
{Tempel} E.,  {Kruuse} M.,  {Kipper} R.,  {Tuvikene} T.,  {Sorce} J.~G.,
  {Stoica} R.~S.,  2018, \mn@doi [\aap] {10.1051/0004-6361/201833217}, \href
  {http://adsabs.harvard.edu/abs/2018A%26A...618A..81T} {618, A81}

\bibitem[\protect\citeauthoryear{{Tully}}{{Tully}}{1988}]{tul88}
{Tully} R.~B.,  1988, {Nearby galaxies catalog}

\bibitem[\protect\citeauthoryear{{Tully} et~al.,}{{Tully} et~al.}{2013}]{tul13}
{Tully} R.~B.,  et~al., 2013, \mn@doi [AJ] {10.1088/0004-6256/146/4/86}, \href
  {http://adsabs.harvard.edu/abs/2013AJ....146...86T} {146, 86}

\bibitem[\protect\citeauthoryear{{Vale} \& {Ostriker}}{{Vale} \&
  {Ostriker}}{2006}]{val06}
{Vale} A.,  {Ostriker} J.~P.,  2006, \mn@doi [\mnras]
  {10.1111/j.1365-2966.2006.10605.x}, \href
  {http://adsabs.harvard.edu/abs/2006MNRAS.371.1173V} {371, 1173}

\bibitem[\protect\citeauthoryear{{Vogelsberger} et~al.,}{{Vogelsberger}
  et~al.}{2014}]{vog14}
{Vogelsberger} M.,  et~al., 2014, \mn@doi [\mnras] {10.1093/mnras/stu1536},
  \href {http://adsabs.harvard.edu/abs/2014MNRAS.444.1518V} {444, 1518}

\bibitem[\protect\citeauthoryear{{Wojtak} et~al.,}{{Wojtak}
  et~al.}{2018}]{woj18}
{Wojtak} R.,  et~al., 2018, \mn@doi [\mnras] {10.1093/mnras/sty2257}, \href
  {http://adsabs.harvard.edu/abs/2018MNRAS.481..324W} {481, 324}

\bibitem[\protect\citeauthoryear{{York}}{{York}}{2000}]{yor00}
{York} D.~G.~S.,  2000, \mn@doi [\aj] {10.1086/301513}, \href
  {http://adsabs.harvard.edu/abs/2000AJ....120.1579Y} {120, 1579}

\bibitem[\protect\citeauthoryear{{de Vaucouleurs}}{{de
  Vaucouleurs}}{1975}]{dev75}
{de Vaucouleurs} G.,  1975, \mn@doi [\apj] {10.1086/154014}, \href
  {http://adsabs.harvard.edu/abs/1975ApJ...202..610D} {202, 610}

\makeatother
\end{thebibliography}

\label{lastpage}
\end{document}